\documentclass[11pt,a4paper]{article}

\pdfoutput=1
\usepackage{jheppub}
\usepackage[T1]{fontenc}
\usepackage[ansinew]{inputenc}
\usepackage[english]{babel}
\usepackage{color}
\usepackage{epsfig, palatino}
\usepackage{pstricks,pst-node,pst-tree}
\usepackage{epic}
\usepackage{mathrsfs}
\usepackage{ae} 
\usepackage{amsmath}
\usepackage{autobreak}
\usepackage{amssymb}
\usepackage{graphicx}
\usepackage{color}
\definecolor{darkblue}{cmyk}{0.9,0.9,0,0}
\usepackage{wasysym}
\usepackage{varioref}
\usepackage{makeidx}
\usepackage{simplewick}
\usepackage{array}
\usepackage{mathrsfs}
\usepackage{feynmf} 
\usepackage{tikz}
\usepackage{dsfont}
\usepackage{slashed}
\usepackage{float}
\usepackage{subcaption}
\usepackage{lscape}
\usepackage{afterpage}

\makeatletter
\DeclareRobustCommand*{\bfseries}{%
  \not@math@alphabet\bfseries\mathbf
  \fontseries\bfdefault\selectfont
  \boldmath
}

\makeatother

\usepackage[textsize=footnotesize]{todonotes}

\usetikzlibrary{decorations.pathmorphing}

\newcommand{\comment}[1]{}

\newcommand{\beq}{\begin{equation}}
\newcommand{\eeq}{\end{equation}}
\newcommand{\beqq}{\begin{equation*}}
\newcommand{\eeqq}{\end{equation*}}
\newcommand\beqa{\begin{eqnarray}}
\newcommand\eeqa{\end{eqnarray}}
\newcommand\beqaa{\begin{eqnarray*}}
\newcommand\eeqaa{\end{eqnarray*}}
\newcommand\bea{\begin{array}}
\newcommand\eea{\end{array}}

\newcommand{\nn}{\nonumber}

\newcommand{\neqa}{\nonumber\end{eqnarray}} 
\newcommand{\la}[1]{\label{#1}}

\renewcommand{\d}{\partial}

\newcommand{\<}{{\langle}}
\renewcommand{\>}{{\rangle}}

\newcommand{\cA}{{\cal A}}

\newcommand{\re}{\relax{\rm I\kern-.18em R}}

\renewcommand{\sp}{p\hspace{-.40em}/}

\newcommand{\phaneq}{\phantom{{}=}}

\def\XXint#1#2#3{{\setbox0=\hbox{$#1{#2#3}{\int}$}
\vcenter{\hbox{$#2#3$}}\kern-.5\wd0}}

\def\[{\left[}
\def\]{\right]}

\def\({\left(}
\def\){\right)}
\def\[{\left[}
\def\]{\right]}

\def\<{\langle}
\def\>{\rangle}

\def\i2{\frac{i}{2}}

\def\cO{{\mathcal O}}

\def\spi{\relax{\rm \pi\kern-0.5em /}}
\def\sA{\relax{\rm A\kern-0.5em /}}
\def\sp{\relax{\rm p\kern-0.5em /}}
\def\sd{\relax{\rm \d\kern-0.5em /}}
\def\sk{\relax{\rm k\kern-0.5em /}}
\def\sn{\relax{\rm n\kern-0.5em /}}
\def\sl{\relax{\rm l\kern-0.5em /}}
\def\sP{\relax{\rm P\kern-0.7em /}}
\def\sBethe{\relax{\rm \Bethe\kern-0.5em /}}

\def\cR{{\cal R}}

\def\cO{{\cal O}}

\def\cW{{\cal W}}

\def\cT{{\cal T}}

\def\2F1{\,_2{\rm F}_1}

\def \Ascr {\cA^{\,^{\kern-0.2em (\infty)}}}

\makeindex

\numberwithin{figure}{section}
\title{An Operator Product Expansion for Form Factors II.\\ Born level}

\author[a]{~~Amit Sever,}
\author[b]{~~Alexander G. Tumanov,}
\author[c]{~~Matthias Wilhelm}

\affiliation[a]{School of Physics and Astronomy, Tel Aviv University, Ramat Aviv 69978, Israel}
\affiliation[b]{Max-Planck-Institut f{\"u}r Physik, Werner-Heisenberg-Institut, 80805 M{\"u}nchen, Germany}
\affiliation[c]{Niels Bohr Institute, 2100 Copenhagen \O{}, Denmark}

 \abstract{Form factors in planar ${\cal N}=4$ Super-Yang-Mills theory admit a type of non-perturbative operator product expansion (OPE), as we have recently shown in \cite{Sever:2020jjx}. This expansion is based on a decomposition of the dual periodic Wilson loop into elementary building blocks: the known pentagon transitions and a new object that we call form factor transition, which encodes the information about the local operator. In this paper, we compute the two-particle form factor transitions for the chiral part of the stress-tensor supermultiplet at Born level; they yield the leading contribution to the OPE. To achieve this, we explicitly construct the Gubser-Klebanov-Polyakov two-particle singlet states. 
The resulting transitions are then used to test the OPE against known perturbative data and to make higher-loop predictions.
}

\preprint{MPP-2021-58}

\begin{document}
\maketitle

\section{Introduction}

One of the most fundamental quantities in QFT are form factors (FFs). They describe overlaps of states created by local operators with $n$-particle asymptotic states. Most of the studies of FFs in $d>2$ QFTs so far have been done in perturbation theory. 

In ${\cal N}=4$ SYM theory in $d=4$, FF could in particular be calculated by generalizing many of the perturbative methods for computing scattering amplitudes, see e.g.\ the recent review \cite{Yang:2019vag}. 
Moreover, integrable structures in FFs have been exploited at strong coupling \cite{Maldacena:2010kp,Gao:2013dza} as well as at weak coupling \cite{Frassek:2015rka}.

In \cite{Sever:2020jjx}, we have put forward a new non-perturbative approach for computing FFs in planar ${\cal N}=4$ SYM theory, called {\it form factor operator product expansion} (FFOPE). It is based on the dual description of planar FFs in terms of certain periodic null polygonal Wilson loops \cite{Alday:2007he,Maldacena:2010kp,
Brandhuber:2010ad,Ben-Israel:2018ckc,Bianchi:2018rrj}, called {\it wrapped Wilson loops}. In this paper, we implement in detail the FFOPE at leading order in perturbation theory for the operators in the chiral half of the stress-tensor supermultiplet and use it to generate new predictions for FFs at all loop orders.
These predictions have already played a crucial role in the perturbative bootstrapping of these form factors \cite{Dixon:2020bbt,PerturbativeBootstrap2}.  

\begin{figure}[t]
\centering
\includegraphics[width=15cm]{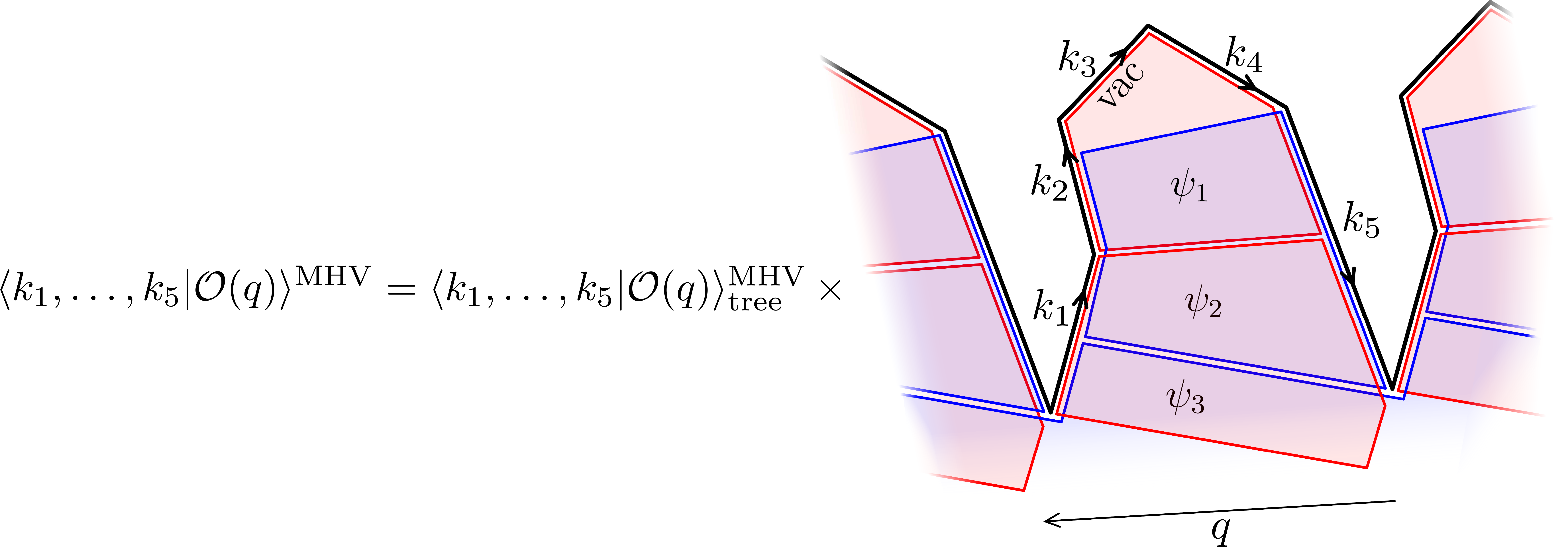}
\caption{\small In the planar limit, an MHV form factor is equal to the expectation value of a wrapped polygonal Wilson loop, multiplied by the tree-level form factor. In the OPE approach, such an $n$-sided wrapped polygon is decomposed into a sequence of pentagons and a two-sided wrapped polygon. Every two consecutive pentagons overlap on a null square and every two consecutive null squares form a pentagon. The last pentagon likewise overlaps with the two-sided wrapped polygon on a null square. Every square that arises from these overlaps shares two of its opposite cusps with the $n$-sided wrapped polygon. For the last (bottom) square, these two cusps coincide with one of the cusps of the two-sided wrapped polygon and its periodic image. Here, this decomposition is illustrated for $n=5$.
}
\label{OPEfig}
\end{figure}

We now review the FFOPE construction of \cite{Sever:2020jjx}, focusing on the elements that enter at leading order in perturbation theory. The starting point of this approach is the dual formulation of FFs in terms of wrapped polygonal Wilson loops \cite{Alday:2007he,Maldacena:2010kp,
Brandhuber:2010ad,Ben-Israel:2018ckc,Bianchi:2018rrj}, as shown in figure~\ref{OPEfig}. These Wilson loops are periodic and are defined in the planar theory with a periodicity constraint that twists the color trace by a spacetime translation, see \cite{Ben-Israel:2018ckc,Cavaglia:2020hdb}. The OPE for these wrapped polygonal Wilson loops mirrors the pentagon operator product expansion (POPE) \cite{Basso:2013vsa, Basso:2013aha,Basso:2014koa,Basso:2014nra,Basso:2014hfa,Basso:2015rta,Basso:2015uxa,Belitsky:2014sla,Belitsky:2014lta,Belitsky:2016vyq} developed for closed polygonal Wilson loops, which are dual to scattering amplitudes. In the FFOPE approach, the wrapped Wilson loop is decomposed into two types of OPE building blocks -- the {\it pentagon transitions}, which correspond to the sequence of closed pentagons in figure \ref{OPEfig}, and the {\it FF transition} that corresponds to the two-sided wrapped polygon at the bottom of the figure. This sequence of pentagon transitions and the FF transition is stitched together by summing over all Gubser-Klebanov-Polyakov (GKP) \cite{Gubser:2002tv} flux-tube excitations that propagate on the squares on which the polygons overlap. 

The OPE limit is the limit in which the edges of the polygon or, equivalently, the momenta of the external gluons become collinear. This kinematical limit is conformally equivalent to stretching the polygon and results in 
a controlled expansion in which the contributions of heavier states are suppressed.

In order to compute planar FFs using the OPE approach, one needs to know the two aforementioned building blocks: the pentagon and the form factor transitions. These transitions are subject to sets of axioms that allow one to bootstrap them at finite 't Hooft coupling, based on the integrability of the GKP flux tube \cite{Basso:2010in}. The pentagon transitions have been extensively studied and bootstrapped in \cite{Basso:2013vsa, Basso:2013aha,Basso:2014koa,Basso:2014nra,Basso:2014hfa,Basso:2015rta,Basso:2015uxa,Belitsky:2014sla,Belitsky:2014lta,Belitsky:2016vyq}. In this paper, we compute the remaining ingredient, the FF transitions, at the leading non-trivial order in perturbation theory. This is the only object that encodes the dependence of the FF on the local operator. We have taken the operator to be in the chiral half of the stress-tensor supermultiplet. The FF transitions we construct are then used to compare the OPE predictions with available perturbative data and generate certain higher loop FF predictions. In \cite{Toappear2}, we present the finite-coupling bootstrap of the FF transitions. 

\begin{figure}[t]
\centering
\includegraphics[width=10cm]{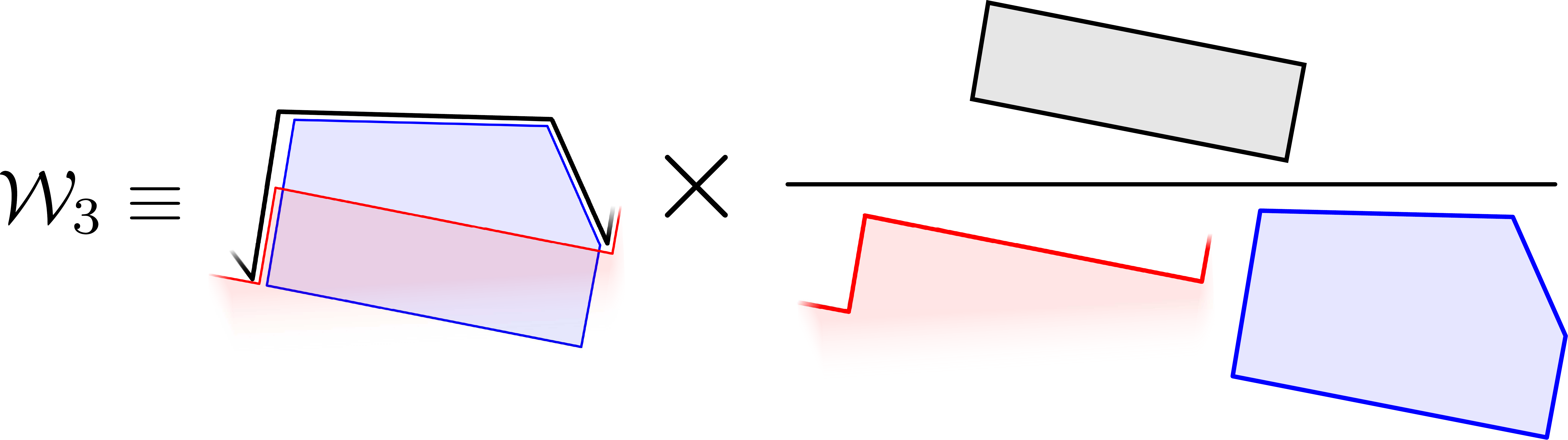}
\caption{A finite conformally invariant ratio is constructed by dividing the three-sided wrapped polygon by the pentagon and the two-sided wrapped polygon and multiplying by the square on which these two objects overlap, $\mathcal{W}_{3}= \frac{\langle W_{3\text{-pt ff}}\rangle\times\langle W_\text{square}\rangle}{\langle W_{2\text{-pt ff}}\rangle\times\langle W_\text{pentagon}\rangle}
$.
}
\label{ratiofig}
\end{figure}

The OPE decomposition for an $n$-particle FF consists of $n-2$ pentagon transitions and a single FF transition. Hence, for the purpose of studying the FF transition, it is sufficient to consider the simplest FF with a non-trivial OPE decomposition. It is the three-gluon FF, for which the OPE decomposition takes the following simple form:
\beq\la{W3}
\mathcal{W}_3= \sum\limits_{\bf a}\int d\textbf{u}\,P_{\bf a}(0|\textbf{u})\,F_{\bar{{\bf a}}}(\bar{\textbf{u}})\,e^{-\tau E(\textbf{u})\,+\,i\sigma p(\textbf{u})}\, .
\eeq
In what follows, we summarize the definitions of the elements entering this equation, while referring the reader to \cite{Sever:2020jjx} for more details. The finite conformal invariant ratio ${\cal W}_3$ is defined in figure~\ref{ratiofig}.
The sum on the r.h.s.\ of (\ref{W3}) is over all GKP flux-tube states, which are parametrized by the number of excitations $N$, their species ${\bf a}=\{a_1,\dots,a_N\}$ and their flux-tube Bethe rapidities ${\bf u}=\{u_1,\ldots,u_N\}$, with $\bar{\bf a} = \{a_N,\ldots,a_1\}$ and $\bar{\textbf{u}}=\{-u_N,\dots,-u_1\}$. The energy $E({\bf u})$ and momentum $p({\bf u})$ of the excitations are known at any value of the 't Hooft coupling \cite{Basso:2010in}, and they are conjugate to the flux-tube time  $\tau$ and space $\sigma$ coordinates respectively. These two independent conformal cross ratios are functions of the space-time momenta $k_1$, $k_2$ and $k_3$ which uniquely parametrize ${\cal W}_3$; they are defined in figure 2 of \cite{Sever:2020jjx}. 
$P_{\bf a}$ are the pentagon transitions. The integration measure is given by
\beq\label{Na}
d\textbf{u} = \mathcal{N}_{\bf a}\,\prod\limits_{i=1}^N \mu_{a_i}(u_i)\,\frac{du_i}{2\pi}\, ,
\eeq
with $\mu_a$ being the single-particle measures and ${\cal N}_{\bf a}$ being a symmetry factor. Lastly, $F_{\bf a}$ are the FF transitions.

In the OPE limit of large $\tau$, which is dual to the near collinear limit of the FF, the expansion (\ref{W3}) is dominated by the lightest contributing excitations. Since the energy of a multi-particle state equals the sum of the energies of the individual particles, one may expect (\ref{W3}) to be dominated by single-particle states. It turns out, however, that FF transitions for the single-particle states are identically zero. 
This is due to the two-sided wrapped polygon being neutral under the $SU(4)_R$ $\cal R$-symmetry and the spacetime $U(1)_\phi$ rotation symmetry in the transverse plane, and therefore being unable to absorb single-particle GKP states, which are all charged under at least one of these symmetries. As a result, the lightest states that contribute to (\ref{W3}) are $SU(4)_R\times U(1)_\phi$ singlet states that consist of two particles \cite{Sever:2020jjx}. 

Each GKP eigenstate corresponds to a certain superposition of scalar, fermion and gluon field strength insertions on an edge of the Wilson loop. 
In perturbation theory, one finds three two-particle singlet states with the same tree-level energy $E=2$, as well as two effective one-particle singlet states. The three two-particle singlet states are superpositions of two scalar ($\phi\bar{\phi}$), two fermion ($\psi\bar{\psi}$) and two gluon ($F\bar{F}$) fields inserted on the edge. In the asymptotic limit where the two fields are taken far apart, only one of these three pairs survives. This surviving pair is used to label the corresponding state. The two-particle singlet states are considerably more complicated than two-particle states build from two  identical fields, and have not been constructed before. One of the main results of this paper is their explicit construction at leading order in perturbation theory. This is accomplished in section \ref{WFWFsection} by diagonalizing the flux-tube transfer matrix in the singlet sector. The other type of excitations, the effective one-particle singlet states $F_{+-}$ and $F_{z\bar z}$, can be shown not to contribute at Born level, see \cite{Sever:2020jjx}.

\begin{figure}[t]
\centering
\includegraphics[width=7.5cm]{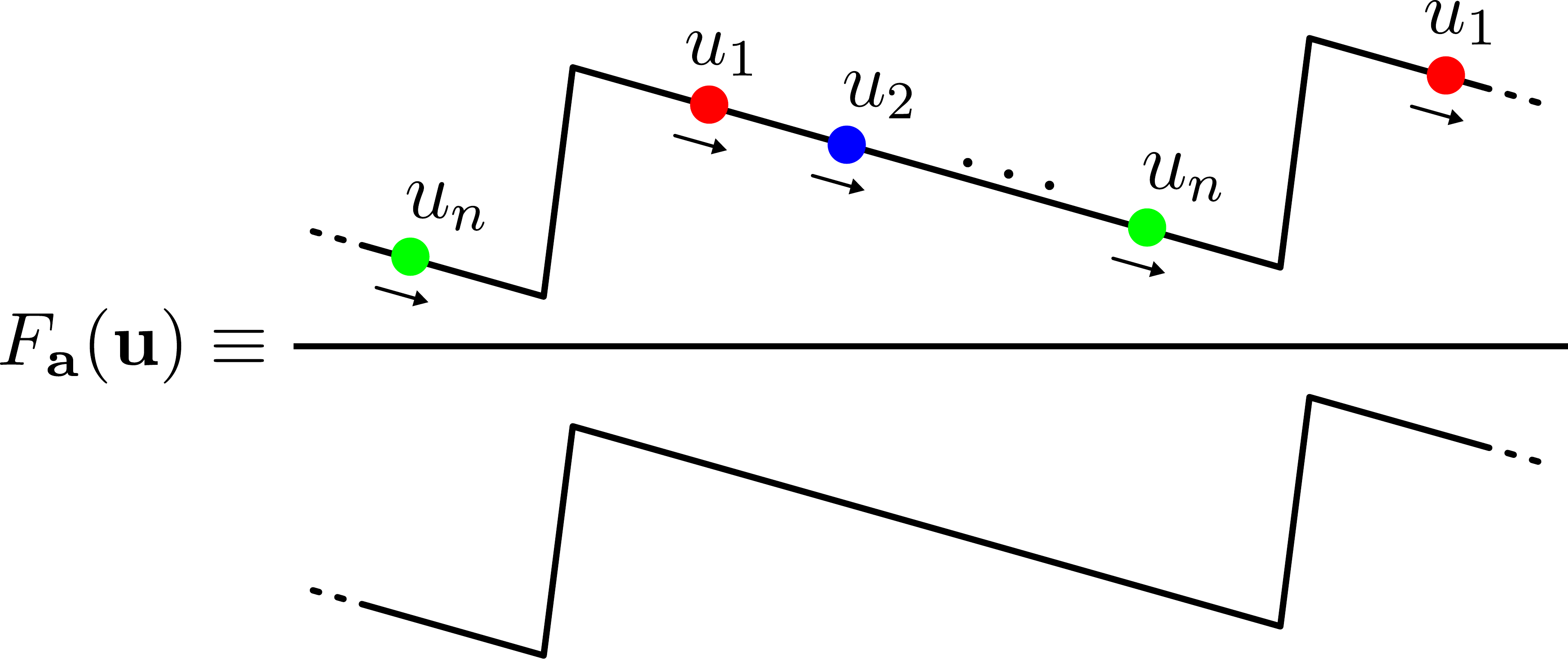}
\caption{The FF transition 
is given by the ratio between the expectation value of the two-sided wrapped polygon with and without GKP excitations inserted on its base. 
}
\label{FFtransdefinition}
\end{figure}

With the explicit states (or equivalently, the dressed Wilson loop operators) at hand, we compute the corresponding FF transitions. This is done in section \ref{FFsec} by evaluating the wrapped polygon with field insertions at leading order in perturbation theory, see figure~\ref{FFtransdefinition}. 
In an upcoming paper \cite{Toappear2}, we will demonstrate how to bootstrap the two-particles FF transitions at finite 't Hooft coupling; 
the Born-level results derived in the present paper will serve as crucial input for this construction.

In section \ref{data}, we use the resulting transitions together with (\ref{Na}) to produce OPE predictions for the three-particle FF at order $e^{-2\tau}$, providing explicit results up to seven-loop order. These match with the independent available data up to two-loop order \cite{Brandhuber:2012vm}, and have already been used for a perturbative bootstrap of the three-point form factor up to seven-loop order \cite{Dixon:2020bbt,PerturbativeBootstrap2}. 

We end this paper with a discussion and future directions in section \ref{Conclusions}, where we also mention a curious observation that allows us to match data for the three-particle FF beyond order $e^{-2\tau}$.

Three appendices provide details on the $SL(2|4)$ symmetry used in the constructions of the two-particle singlet wave functions (appendix \ref{SL24appendix}), square and pentagon transitions (appendix \ref{SWFappendix}) as well as checks of the wave function using the pentagon transitions (appendix~\ref{PTappendix}). Ancillary files attached to this publication include explicit expressions for the OPE predictions for the three-point form factor up to seven-loop order (\texttt{OPE-predictions.txt}), as well as the implementations of the wave functions, their checks and construction of the FF and pentagon transitions from them (\texttt{Born level form factor OPE.nb}).

\section{The Two-Particle Singlet States}\label{WFWFsection}

In this section, we construct the three lightest two-particle singlet states at Born level by diagonalizing the corresponding transfer matrices.

\subsection*{Flux-tube states and symmetry}

In the OPE picture, the flux-tube vacuum is represented by the square Wilson loop.
The flux-tube time $\tau$ is conjugate to a conformal symmetry of the square which moves points from its bottom towards the top. This conformal generator, called twist, 
leads to a sort of operator-state correspondence between flux-tube excitations on top of the vacuum and adjoint operators inserted along the bottom or top edges of the square. These excitations form a Fock space of gaped multi-particle states. To study them, it is very useful to first understand the symmetries of the problem.

The twist conformal symmetry of the square 
commutes with the $SL(2,\mathbb{R})_\sigma$ conformal symmetry of the infinite null lines along the bottom and top edges of the square. This bosonic symmetry is extended to an $SL(2|4)\subset PSU(2,2|4)$ super-conformal symmetry of ${\cal N}=4$ SYM theory, see appendix~\ref{SL24appendix} for details. The right and left edges of the square brake it down to a non-compact $U(1)_\sigma\times SU(4)_R$ subgroup \cite{Alday:2010ku}. 
As a result, 
the flux-tube Hamiltonian \cite{Belitsky:2011nn,Sever:2012qp,Belitsky:2004cz,Basso:2013aha} only acts on $SL(2,\mathbb{R})_\sigma$ conformal cross ratios built out of the coordinates along the bottom edge of the left and right edges, together with those of the operator insertion points along the bottom and top edges. In the planar limit, this symmetry is further extended to a full Yangian symmetry, for which the right and left edges serve as integrable boundary conditions. 

\subsection*{Born-level flux-tube dynamics}
In this paper, we are only considering Born-level flux-tube states. Since the tree-level Hamiltonian is highly degenerate, these states are constructed by diagonalizing the one-loop Hamiltonian. 
There are few important simplifications that come about at this order.
Firstly, at Born-level the GKP vacuum has an additional $SL(2,\mathbb{R})_\tau$ symmetry, of which the Hamiltonian is one of the generators \cite{Alday:2010ku}.\footnote{This is because the creation of excitations by the $SL(2,\mathbb{R})_\tau$ generators 
costs powers of the coupling and is not seen at Born level.
} As a result, the flux-tube states are organized into $SL(2,\mathbb{R})_\tau$ primaries and descendants. 
Secondly, an $n$-particle state has a simple operator representation in terms of a superposition of $n$ adjoint field insertions on the Wilson line. 
In the remainder of this section, we will explicitly construct this superposition for the two-particle singlet states.

Single-field $SL(2,\mathbb{R})_\tau$ primaries of bare twist $1, 2,\dots$ correspond to different Born-level single-particle excitations, see \cite{Gaiotto:2011dt} for details. Among these, the twist-1 excitations are the lightest. They correspond to the following types of field insertions:  
two gauge fields $F$ and $\bar F$, eight fermions $\psi_A$ and $\bar\psi^A$, as well as six scalars $\phi_{AB}=-\phi_{BA}$, where $A=1,2,3,4$ is an $SU(4)_R$ index. They carry $U(1)_\phi$ charge $\pm 1$, $\pm \frac{1}{2}$ and $0$ respectively, where the $U(1)_\phi$ symmetry corresponds to rotations in the plane transverse to the square. Together, they form a single multiplet of $SL(2|4)$, which can be conveniently represented in terms of a (type of) superfield in the following way:
\begin{align}\label{supermultipletdef}
\Phi(x,\theta^A)=\,& F(x) + \theta^A\psi_A(x) + \frac{1}{2!}\theta^A\theta^B\phi_{AB}(x)\nonumber\\
&+ \frac{1}{3!}\varepsilon_{ABCD}\theta^A\theta^B\theta^C\bar{\psi}^D(x) + \frac{1}{4!}\varepsilon_{ABCD}\theta^A\theta^B\theta^C\theta^D\bar{F}(x)\, ,
\end{align}
where $\theta^{A=1,2,3,4}$ are a set of Gra{\ss}mann bookkeeping parameters that transform in the fundamental representation of $SU(4)_R\subset SL(2|4)$.

This superfield is inserted along the bottom edge of the square. We choose this edge to extend along the $x^-$ direction and parametrize points on it as $x^-(\sigma)$, where $\sigma\in(-\infty,\infty)$ parametrizes the $SL(2,\mathbb{R})_\sigma$ conformal symmetry that preserves this edge, see \cite{Basso:2013aha} for details.
In what follows, we will drop the superscript $-$ of $x^-$. A state with $n$ excitations inserted at positions $x(\sigma_1),\dots,x(\sigma_u)$ corresponds to the dressed Wilson line operator
\beq\label{statedef}
|\boldsymbol{\sigma},\boldsymbol{\theta}\rangle =\star\,\left(x'(\sigma_1)\right)^{\hat{\bf s}_1}\Phi(x(\sigma_1),\theta_1)\,\star\,\ldots\,\star\,\left(x'(\sigma_n)\right)^{\hat{\bf s}_n}\Phi(x(\sigma_n),\theta_n)\,\star\, ,\nonumber
\eeq
where $\boldsymbol{\sigma}$ and $\boldsymbol{\theta}$ are shorthand notations for $\{\sigma_1,\ldots,\sigma_n\}$ and $\{\theta_1,\ldots,\theta_n\}$ respectively, and $\,\star\,$ stands for the corresponding section of the Wilson line.
The factors $\left(x'(\sigma_i)\right)^{\hat{\bf s}_i}$ are included in this definition in order to remove the $SL(2,\mathbb{R})_\sigma$ conformal weight, with $\hat{\bf s}_i=(1+|2-\theta_i\partial_{\theta_i}|)/2$ measuring the conformal spin of the corresponding field. 

A general $n$-particle flux-tube state is decomposed in this basis as 
\beq\label{wfdeff}
|\Psi\rangle = \int\limits_{\sigma_1<\ldots<\sigma_n} d^n\boldsymbol{\sigma}\,d^{4n}\boldsymbol{\theta}\,\Psi(\boldsymbol{\sigma},\boldsymbol{\theta})|\boldsymbol{\sigma},\boldsymbol{\theta}\rangle\, ,
\eeq
where $\Psi(\boldsymbol{\sigma},\boldsymbol{\theta})$ is the 
flux-tube wave function. The integration is only performed over the domain $-\infty<\sigma_1<\ldots<\sigma_n<+\infty$ to prevent overcounting.

We are interested in constructing the two-particle singlet states that diagonalize the flux-tube Hamiltonian.  
As a warm-up, we first consider the much simpler case of a single-particle state. 

\subsection*{Single-particle states}
The flux-tube Hamiltonian is an integral operator that mixes fields inserted at different positions on the edge. For a single-particle state, it takes a particularly simple form \cite{Belitsky:2011nn,Belitsky:2005qn}
\beq\la{H1p}
\mathcal{H}\cdot \Psi(\sigma,\theta)=2\int\limits_{-\infty}^\infty \frac{dt}{\sinh|t|}\[e^{-|t|}\Psi(\sigma,\theta)-e^{|(1-\frac{1}{2}\theta\d_\theta)t|}
\Psi(\sigma+t,\theta)\]\,.
\eeq
Due to the translation symmetry of $\mathcal{H}$ in (\ref{H1p}), it is trivially diagonalized in Fourier space
\beq\la{HPsi}
\mathcal{H}\cdot \Psi_{v,a}(\sigma,\theta) = E_{\frac{1}{2}(1+|2-a|)}^{(1)}(v)\, \Psi_{v,a}(\sigma,\theta)\, ,\qquad\Psi_{v,a}(\sigma,\theta)
= e^{2iv\sigma}\theta^a\, ,
\eeq
where the $SU(4)_R$ index is suppressed and $a=0,1,2,3,4$, with $\theta^1=\theta_A$, $\theta^2=\epsilon_{ABCD}\theta^A\theta^B$, etc.
The corresponding energies are
\beq\la{onelE}
E_{s}^{(1)}(v) =2\[ \psi(s+iv)+\psi(s-iv)-2\,\psi(1)\]\, ,
\eeq
where $\psi(u) = \frac{d}{dv}\,{\rm log}\,\Gamma(v)$ is the digamma function. 

For more than one particle, the flux-tube Hamiltonian is a more complicated integral operator that is difficult to diagonalize directly. Due to the integrability of the GKP flux tube, however, the flux-tube energy is just one of an infinite set of commuting conserved charges. Hence, instead of diagonalizing the Hamiltonian directly, we may choose to diagonalize any other set of conserved charges that is sufficient to fix the eigenstates uniquely. One such set of conserved charges is generated by the transfer matrices in the fundamental or anti-fundamental representations. Unlike the Hamiltonian, the transfer matrix in the fundamental representation is a differential operator, and thus it is easier to diagonalize. For instance, in the one-particle case considered above, the one-particle transfer matrices in the fundamental and anti-fundamental representations are 
combinations of the 
differential operators $\d_\sigma$ and $\theta\d_\theta$:
\beq\la{T1}
T_1(u) =(u + i) - \frac{i}{2}\left(\partial_\sigma+\theta \partial_\theta\right)\,,
\eeq 
where $u$ is the spectral parameter. The anti-fundamental transfer matrix is obtained from the fundamental one by changing $\theta\d_\theta\to4-\theta\d_\theta$:
\beq\la{T1b}
\bar{T}_1(u) =\left.T_1(u)\right|_{\theta\d_\theta\to4-\theta\d_\theta}=(u - i) - \frac{i}{2}\left(\partial_\sigma-\theta \partial_\theta\right)\,.
\eeq 
The eigenfunctions of the transfer matrices (\ref{T1}) and (\ref{T1b}) are fixed by demanding that the wave function does not grow at infinity. It is therefore apparent that the one-particle state can be obtained by simultaneously diagonalizing the operators $\d_\sigma$ and $\theta\d_\theta$.

As a preparation for the diagonalization in the two-particle case, we will now construct the two-particle transfer matrices in the fundamental and anti-fundamental representations.

\subsection*{The transfer matrix}

The transfer matrix is an operator that acts on the GKP states. It can be thought of 
as representing the propagation of an (unphysical) auxiliary particle between a certain state on the right edge of the square to a certain state on the left edge. Along the way, this auxiliary particle scatters off the physical GKP excitations:
\beq\la{Tgen}
T_n(u)=\frac{s_L\cdot\cR_1(u)\cdot\ldots\cdot\cR_n(u)\cdot s_R}{ s_L\cdot s_R}\, .
\eeq
Here, $s_L, s_R$ are the left and right boundary states of the auxiliary particle and the dot 
represents the $SL(2|4)$-invariant product in the auxiliary space. The R-matrix, $\cR_j$, describes the scattering between the auxiliary and the $j$'th physical particle. The most important property of the R-matrix is that it obeys a triangular relation known as Yang-Baxter equation. 

The auxiliary particle can transform in different representations of $SL(2|4)$. Each representation results in a different transfer matrix, all commuting with each other for any values of the spectral parameters. The simplest representations one may pick are the fundamental and anti-fundamental (non-unitary) representations of $SL(2|4)$, for which the transfer matrix takes the form
\beq\la{Rform}
\cR(u)\ =\  u\,{\mathbb I}^\text{(aux)}\otimes{\mathbb I}^\text{(phys)}+\frac{i}{2}\sum_a{\mathbb J}^\text{(aux)}_a\otimes{\mathbb J}^\text{(phys)}_a\, ,
\eeq
where the spectral parameter $u$ represents the momentum of the auxiliary particle and ${\mathbb J}_a$ are the $SL(2|4)$ generators. The generators ${\mathbb J}^\text{(aux)}$ act on the auxiliary space and ${\mathbb J}^\text{(phys)}$ act on the physical space. The combination of generators in (\ref{Rform}) is invariant under simultaneous $SL(2|4)$ transformations of the physical and auxiliary spaces. It is the same form that also appears in the quadratic Casimir (\ref{Cas}), the only difference being that in the case of the $R$-matrix the two generators in ${\mathbb J}^\text{(aux)}_a\otimes\mathbb J^\text{(phys)}_a$ act on different spaces. The auxiliary space is represented by a 6-dimensional vector transforming in the fundamental/anti-fundamental representations while the physical space is represented by the super-field insertion $\left(x'(\sigma)\right)^{\hat{\bf s}}\Phi(x(\sigma),\theta)$, which forms a non-compact unitary representation. 

The $SL(2|4)$ generators can be divided into bosonic and fermionic ones. The bosonic (Gra{\ss}mann-even) ones are the $SL(2,\mathbb{R})\subset SL(2|4)$ generators $\{L_\pm,L_0\}$, the $SU(4)_R\subset SL(2|4)$ generators ${T_A}^B$, and another $SL(2,\mathbb{R})\times SU(4)_R$ singlet bosonic generator $B$ that is needed to close the algebra. The fermionic (Gra{\ss}mann-odd) super-conformal generators are $\{V_{\pm,A},W_\pm^A\}$. 
Their commutation relations as well as the explicit representation on the auxiliary and physical spaces are given in appendix \ref{SL24appendix}. In terms of these generators, the $R$-matrix takes the form
\begin{align}\la{RRbar}
\mathbb{R}(u) = u\,\mathbb{I}_{2\times 2} + i&\left(2\,\mathcal{L}_{0}\,\mathbb{L}_{0} - \mathcal{L}_{+}\,\mathbb{L}_{-} - \mathcal{L}_{-}\,\mathbb{L}_{+} + 4\,\mathcal{B}\,\mathbb{B} - \mathcal{T}_{A}{}^B\,\mathbb{T}_{B}{}^A\right.\\
&\left.+\,\mathcal{V}_{+,A}\,\mathbb{W}_{-}^{A} - \mathcal{V}_{-,A}\,\mathbb{W}_{+}^{A} + \mathcal{W}_{+}^A\,\mathbb{V}_{-,A} - \mathcal{W}_{-}^A\,\mathbb{V}_{+,A}\right)\, ,\nonumber
\end{align}
where we used calligraphic font to represent the auxiliary generators and double-strike font for the physical ones.

The boundary state in the auxiliary space, $s_L$ ($s_R$), is a primary state of $SL(2,\mathbb{R})_\sigma$ with respect to the position of the left (right) edge. In other words, in the representation where ${\mathbb L}_-$ (${\mathbb L}_+$) leaves the position of the left (right) edge invariant, ${\cal L}_-$ (${\cal L}_+$) annihilates $s_L$ ($s_R$). These states are called ``small solutions'' \cite{Sever:2012qp}. When lifted to $SL(2|4)$, $s_L$ is promoted to an element of the $\bar{\textbf{6}}$ representation, while $s_R$ becomes an element of $\textbf{6}$ of $SL(2|4)$. This promotion is achieved, for example, by additionally demanding $s_R$ to be annihilated by ${\mathcal V}_\pm$ and $s_L$ to be annihilated by $\bar{{\mathcal W}}_\pm$. This ensures that the resulting transfer matrices commute. Precise expressions for the small solutions we are using to obtain the results of this paper are given in appendix~\ref{SL24appendix}.

Inserting the representations of the generators and the small solutions from appendix~\ref{SL24appendix} into (\ref{RRbar}) and (\ref{Tgen}), we find for $n=1$ the one-particle transfer matrices given in \eqref{T1}-\eqref{T1b}. Using integration by parts, one can then express the transfer matrix as a differential operator acting on wave functions $\Psi(\sigma,\theta)$, instead of on the line operators $|\sigma,\theta\>$: 
\beq
\int d\sigma\,d^4\theta\[\mathcal{T}_1(u)\,\Psi(\sigma,\theta)\]|\sigma,\theta\rangle\equiv\int d\sigma\,d^4\theta\,\Psi(\sigma,\theta)\[T_1(u)|\sigma,\theta\rangle\]\, .
\eeq
This procedure has the simple effect of flipping the signs of $\d_\sigma$ and $u$ as well as replacing $\theta\d_\theta\to(4-\theta\d_\theta)$, which boils down to an overall sign flip: 
\beq\la{cT1}
\cT_1(u) =(u - i) + \frac{i}{2}\left(\partial_\sigma+\theta \partial_\theta\right)\, .
\eeq 
As previously mentioned, the wave functions $\Psi_{v,a}(\sigma,\theta)$ in (\ref{HPsi}) diagonalize the transfer matrix in addition to the Hamiltonian, 
\begin{equation}
 \cT_1(u)\Psi_{v,a}(\sigma,\theta)=\lambda(u|v,a)\Psi_{v,a}(\sigma,\theta)\, ,
\end{equation}
with eigenvalues
\beq
\lambda(u|v,a)=(u-v)-i\left(1-a/2\right)\, .
\eeq
Finally, the anti-fundamental counterpart of $\cT_1(u)$ can likewise be obtained by the $\theta\d_\theta\to(4-\theta\d_\theta)$ replacement,
\beq\la{cT1b}
\bar{\cT}_1(u) = (u + i) + \frac{i}{2}\left(\partial_\sigma-\theta \partial_\theta\right)\, .
\eeq

\subsection*{The two-particle transfer matrix}

We now generalize the construction of the transfer matrix to states that consist of two flux-tube excitations. Consider first the limit in which two excitations are far apart from each other. This limit is particularly simple because the flux-tube spectrum is gapped and hence the two particles decouple from each other. As a result, the energy of the two-particle state is given by the sum of the two one-particle energies and the eigenvalue of the transfer matrix is the product of the two one-particle eigenvalues. We therefore know a priori the exact spectrum of these operators and the asymptotic form of the corresponding wave-functions:
\beq
E(\textbf{v},\textbf{a})=E_{s_1}(v_1)+E_{s_2}(v_2)\, ,\qquad \lambda(u|\textbf{v},\textbf{a})=\lambda(u|v_1,a_1)\times\lambda(u|v_2,a_2)\, ,
\eeq
where $\textbf{v}=\{v_1,v_2\}$, $\textbf{a}=\{a_1,a_2\}$ and $s_i=(1+|2-a_i|)/2$ is the conformal spin of the excitation. We still need to find the corresponding eigen(wave)functions, though, which we denote by $\Psi(\textbf{v},\textbf{a}|\boldsymbol{\sigma},\boldsymbol{\theta})$.  

The full two-particle transfer matrix (\ref{Tgen}) takes the form 
\beq\la{Tstates}
T_2(u)=T_1^{(1)}(u) \,T_1^{(2)}(u)-e^{\sigma_1-\sigma_2}\,\theta_1\partial_{\theta_2}\,\bar M^{(1)}(-\tfrac{i}{2})M^{(2)}(\tfrac{i}{2})-e^{2(\sigma_1-\sigma_2)}\bar D^{(1)}(-\tfrac{i}{2})D^{(2)}(\tfrac{i}{2})\, ,
\eeq
where $T_1^{(i)}(u)$ is the one-particle transfer matrix (\ref{T1}) acting on the $i$'th site. 
The differential operators $D^{(i)}(u)$ and $M^{(i)}(u)$ are given by%
\footnote{The necessity to distinguish cases follows from the slightly unusual definition of the superfield \eqref{supermultipletdef}.}
\beq\la{DMi}
D^{(i)}(u)=\left\{\begin{array}{lcl}T_1^{(i)}(u)&&a_i\le 2\, ,\\ \bar T_1^{(i)}(u)&&a_i>2\, ,\end{array}\right.
\qquad M^{(i)}(u)=\left\{\begin{array}{lcl}T_1^{(i)}(u)&&a_i\le 2\, ,\\ -\,i&&a_i>2\, ,\end{array}\right.
\eeq
while $\bar D^{(i)}$ and $\bar M^{(i)}$ are related to these by replacing $a_i\to4-a_i$ in (\ref{DMi}).

One can check explicitly that two transfer matrices \eqref{Tstates} commute for any values of the spectral parameters,
\begin{align}
\left[T_2(u),T_2(v)\right] = 0\, ,
\end{align}
which implies that they can be diagonalized simultaneously. As in the single-particle case, the two-particle states are uniquely fixed by diagonalizing the fundamental transfer matrix $T_2(u)$ with the appropriate boundary conditions. However, we have found it useful to consider the anti-fundamental transfer matrix $\bar T_2(u)$ in addition to the fundamental one, since, depending on the component, different equations for either the fundamental or anti-fundamental transfer matrix are easier to solve. The anti-fundamental transfer matrix $\bar T_2(u)$ also commutes with $T_2(v)$ and is obtained from it by flipping $\theta_i\d_{\theta_i}\to 4-\theta_{i}\d_{\theta_{i}}$, and $\theta_1\d_{\theta_2}\to-\,\theta_2\d_{\theta_1}$.

Next, we use integration by parts to obtain the representation of the transfer matrix on wave functions $\mathcal{T}$ from its representation on states $T$ (\ref{Tstates}). It is fixed by demanding that for any wave function that does not grow at infinity, the following relation holds
\beq\la{Tstowf}
\int\limits_{\sigma_1<\sigma_2}\!\!\!\!\! d^2\boldsymbol{\sigma}\,d^{8}\boldsymbol{\theta}\ \Psi(\boldsymbol{\sigma},\boldsymbol{\theta}|\textbf{u},\textbf{a})\[T_2(u)|\boldsymbol{\sigma},\boldsymbol{\theta}\rangle\]=\!\!\!\int\limits_{\sigma_1<\sigma_2}\!\!\!\!\! d^2\boldsymbol{\sigma}\,d^{8}\boldsymbol{\theta}\ \[\mathcal{T}_2(u)\,\Psi(\boldsymbol{\sigma},\boldsymbol{\theta}|\textbf{u},\textbf{a})\]|\boldsymbol{\sigma},\boldsymbol{\theta}\rangle\, .
\eeq
Integration by parts has a similar effect as in the one-particle case. 
For two particles, however, the integration domain in (\ref{Tstowf})
has a new boundary at $\sigma_1 = \sigma_2$. 
As a result of this, integration by parts leads to a new boundary term $\delta {\cal T}_2$ that is localized at the point where the two particles collide. It is proportional to $\delta(\sigma_1-\sigma_2+0^+)$ and its derivative, and should be understood in a distributional sense -- as an operator acting on a smooth functions under integration. Here, the $0^+$ in the argument of the $\delta$-function is added to ensure that its support lies inside of the region of integration in (\ref{Tstowf}). All the $\delta$-functions in this paper should be understood in this way. For ease of presentation, we will not be writing these $0^+$'s explicitly from this point on. 

In total, we find
\begin{align}\label{Twf2}
{\cal T}_2(u) =\,\,& {\cal T}_1^{(1)}(u) {\cal T}_1^{(2)}(u)+e^{\sigma_1-\sigma_2}\,\theta_1\partial_{\theta_2}\,{\cal M}^{(1)}(\tfrac{i}{2})\bar{\cal M}^{(2)}(-\tfrac{i}{2})-e^{2(\sigma_1-\sigma_2)}{\cal D}^{(1)}(\tfrac{i}{2})\bar{\cal D}^{(2)}(-\tfrac{i}{2})\nn\\
&+ \delta {\cal T}_2\, ,
\end{align}
where $\mathcal{T}_1^{(i)}(u)$ is the one-particle wave-function transfer matrix (\ref{cT1}), while ${\cal D}^{(i)}(u)$ and ${\cal M}^{(1)}(u)$ are the wave-function analogs of $D^{(i)}(u)$ and $M^{(1)}(u)$ defined in (\ref{DMi})\, ,
\beq\la{cDMi}
\mathcal{D}^{(i)}(u)=\left\{\begin{array}{lcl}\mathcal{T}_1^{(i)}(u)&&a_i< 2\, ,\\ \bar{\mathcal{T}}_1^{(i)}(u)&&a_i\ge 2\, ,\end{array}\right. \qquad \mathcal{M}^{(i)}(u)=\left\{\begin{array}{lcl}\mathcal{T}_1^{(i)}(u)&&a_i< 2\, ,\\ -\,i&&a_i\ge 2\, ,\end{array}\right.
\eeq
while $\bar{\mathcal{D}}^{(i)}$ and $\bar{ \mathcal{M}}^{(i)}$ can be obtained by interchanging $a_i\to4-a_i$ in (\ref{cDMi}).
The boundary term is given by
\begin{align}
\delta{\cal T}_2 = \frac{1}{2}\,\delta{\cal J}_2 - \frac{1}{2}\,\theta_1\partial_{\theta_2}\,\delta{\cal K}_2\, ,
\end{align}
for which
\begin{align}
 \delta{\cal J}_2
 =\delta(\sigma_1-\sigma_2)\times\begin{cases}
   \theta_2\partial_{\theta_2}-2 &\quad\text{for}\quad a_1<2 \,\&\, a_2\leq2\, ,\\
   0&\quad\text{for}\quad a_1<2\,\&\,a_2>2\, ,\\
   \theta_2\partial_{\theta_2}-\theta_1\partial_{\theta_1}&\quad\text{for}\quad a_1\geq2\,\&\,a_2\leq2\, ,\\
   2-\theta_1\partial_{\theta_1}
   &\quad\text{for}\quad a_1\geq2\,\&\,a_2>2\, ,
  \end{cases}
\end{align}
and
\begin{align}
 \delta{\cal K}_2
 =\begin{cases}
   \delta(\sigma_1-\sigma_2) & \text{ for }a_1<2 \,\&\, a_2\leq2\, ,\\
      \frac{1}{2}\left[\delta'(\sigma_1-\sigma_2)+\delta(\sigma_1-\sigma_2)\left(1+\partial_{\sigma_1} + \theta_1\partial_{\theta_1} - \partial_{\sigma_2} - \theta_2\partial_{\theta_2}\right)\right]
  \!\!\!\! & \text{ for }a_1<2 \,\&\, a_2>2\, ,\\
   0 &\text{ for }a_1\geq2 \,\&\, a_2\leq2\, ,\\
   -\,\delta(\sigma_1-\sigma_2)
   &\text{ for }a_1\geq2 \,\&\, a_2>2\, .
  \end{cases}
\end{align}

The wave-function transfer matrices are subject to a multiplication rule in which the squares of the boundary terms are removed, as the squares of boundary terms are not generated through integration by parts:
\begin{align}\label{Dcomm}
\mathcal{T}_2(u)\cdot\mathcal{T}_2(v) \equiv {\cal T}_2(u)\,{\cal T}_2(v)- (\delta{\cal T}_2 )^2\, .
\end{align}
With this multiplication rule, one can check that indeed $\mathcal{T}_2(u)\cdot\mathcal{T}_2(v) -\mathcal{T}_2(v)\cdot\mathcal{T}_2(u) =0$ for any $u$ and $v$. Lastly, the anti-fundamental transfer $\bar{\cal T}_2(u)$ matrix is obtained from ${\cal T}_2(u)$ by flipping $\theta_i\d_{\theta_i}\to 4-\theta_{i}\d_{\theta_{i}}$, and $\theta_1\d_{\theta_2}\to-\theta_2\d_{\theta_1}$. It also commutes with ${\cal T}_2(v)$ with the appropriate multiplication rule. 

\subsection*{The two-particle singlet states}
\label{wfsectionexplicit}

We now construct the two-particle singlet flux-tube eigenstates of bare twist 2 by diagonalizing the transfer matrix (\ref{Twf2}) in the singlet sector. This sector 
is spanned by all combinations of two conjugate fields in (\ref{supermultipletdef}), each having bare twist equal to 1. These are two conjugate gauge fields, two conjugate fermions, and two conjugate scalars inserted at two independent positions along the Wilson line, $\sigma_2>\sigma_1$. Indeed, all of these combinations are $SU(4)_R$ and $U(1)_\phi$ singlets. In terms of the bookkeeping parameters $\theta^A$ in (\ref{wfdeff}), the most general wave function in this sector takes the form
\begin{align}\label{wfansatz}
\Psi_\text{singlet}&(\sigma_1,\sigma_2;\theta_1,\theta_2) = \Psi^{\bar{F}F}(\sigma_1,\sigma_2)\, \theta_1^{\vphantom{[}1}\theta_1^{\vphantom{[}2}\theta_1^{\vphantom{[}3}\theta_1^{\vphantom{[}4} + 4\,\Psi^{\bar{\psi}\psi}(\sigma_1,\sigma_2)\, \theta_1^{[1}\theta_1^{\vphantom{[}2}\theta_1^{\vphantom{[}3}\theta_2^{4]}\\
&+ 6\,\Psi^{\phi\bar{\phi}}(\sigma_1,\sigma_2)\, \theta_1^{[1}\theta_1^{\vphantom{[}2}\theta_2^{\vphantom{[}3}\theta_2^{4]} + 4\,\Psi^{\psi\bar{\psi}}(\sigma_1,\sigma_2)\, \theta_1^{[1}\theta_2^{\vphantom{[}2}\theta_2^{\vphantom{[}3}\theta_2^{4]} + \Psi^{F\bar{F}}(\sigma_1,\sigma_2)\, \theta_2^{\vphantom{[}1}\theta_2^{\vphantom{[}2}\theta_2^{\vphantom{[}3}\theta_2^{\vphantom{[}4}\, ,\nonumber
\end{align}
where
\beq
\theta_i^{[1}\theta_j^{\vphantom{[}2}\theta_k^{\vphantom{[}3}\theta_l^{4]}\equiv\frac{1}{4!}\epsilon_{ABCD}\theta_i^A\theta_j^B\theta_k^C\theta_l^D\, .
\eeq

Recall that the transfer matrix (\ref{Twf2}) consists of a regular bulk part and a singular boundary part. Hence, the space of wave functions they act on must have the same structure. 
More specifically, one finds
\begin{align}\label{wf2part}
\Psi(\sigma_1,\sigma_2;\theta_1,\theta_2) &\equiv \psi(\sigma_1,\sigma_2;\theta_1,\theta_2)+ \delta\psi(\sigma_1,\sigma_2;\theta_1,\theta_2)\, ,
\end{align}
where $\psi$ is the piece that is regular at $\sigma_1=\sigma_2$, and $\delta\psi$ is the piece that is proportional to $\delta(\sigma_1-\sigma_2)$ and its derivatives. The action of the transfer matrix (\ref{Twf2}) on $\Psi$ can be deduced from the integration-by-parts procedure and takes the form%
\footnote{For the rest of the paper, we will suppress the $\sigma$ and $\theta$ arguments of the wave functions to simplify the notation.}
\begin{align}
&\mathcal{T}_2(u)\cdot\Psi \equiv \mathcal{T}_2(u)\,\Psi - \delta \mathcal{T}_2\,\delta\psi\, .
\end{align}
Diagonalizing the transfer matrix in this sector boils down to solving the following eigenvalue equations,
\begin{equation}
 \label{Fulleq}
\begin{aligned}
&\mathcal{T}_2(u)\cdot\Psi_{\Phi\bar{\Phi}} = (u-u_1-i\xi_\Phi)\,(u-u_2+i\xi_\Phi)\,\Psi_{\Phi\bar{\Phi}}\, ,\\
&\bar{\mathcal{T}}_2(u)\cdot\Psi_{\Phi\bar{\Phi}} = (u-u_1+i\xi_\Phi)\,(u-u_2-i\xi_\Phi)\,\Psi_{\Phi\bar{\Phi}}\, ,
\end{aligned}
\end{equation}
where the eigenvalues are fixed by the asymptotic behavior of the solution in the $\sigma_1\ll\sigma_2$ limit. The subscript ${\Phi\bar{\Phi}}$ is equivalent to the particle-species index $\textbf{a} = {a_1,a_2}$ from  (\ref{Tstowf}) adapted to singlet states, for which the two asymptotic particles are conjugate to each other. It can take three different values, ${\Phi\bar{\Phi}}\in\{\phi\bar{\phi},\psi\bar{\psi},F\bar{F}\}$. For instance, ${\Phi\bar{\Phi}}=F\bar{F}$ corresponds to a wave function that describes $F$ with rapidity $u_1$ and $\bar{F}$ with rapidity $u_2$. Note that the subscript ${\Phi\bar{\Phi}}$ labels the wave function as a whole, while the superscript from (\ref{wfansatz}) labels its components.
For convenience, we have assigned the following numerical values $\xi_\Phi$ to the wave functions $\Psi_{\Phi\bar{\Phi}}$ as well as the corresponding states:
\begin{align}
\xi_{\bar{F}} = -\,1\, ,\quad\xi_{\bar{\psi}} = -\,\frac{1}{2}\, ,\quad \xi_{\phi} = 0\, ,\quad\xi_{\psi} = \frac{1}{2}\, ,\quad\xi_{F} = 1\, ,
\end{align}
where the negative values for conjugated fields have been added for use in (\ref{2point35}) and appendices \ref{SWFappendix} and \ref{PTappendix}.

The eigenvalue equations (\ref{Fulleq}) can be separated into the non-singular bulk part and the singular boundary part,
\begin{equation}\label{FulleqRegular}
 \begin{aligned}
&\left[\mathcal{T}_2(u)-\delta\mathcal{T}_2-(u-u_1-i\xi_\Phi)\,(u-u_2+i\xi_\Phi)\right]\psi_{\Phi\bar{\Phi}} = 0\,,\\
&\left[\bar{\mathcal{T}}_2(u)-\delta\bar{\mathcal{T}}_2-(u-u_1+i\xi_\Phi)\,(u-u_2-i\xi_\Phi)\right]\psi_{\Phi\bar{\Phi}} = 0\,,
\end{aligned}
\end{equation}
and
\begin{equation}\label{FulleqSingular}
 \begin{aligned}
&\left[\mathcal{T}_2(u)-\delta\mathcal{T}_2-(u-u_1-i\xi_\Phi)\,(u-u_2+i\xi_\Phi)\right]\delta\psi_{\Phi\bar{\Phi}} = -\,\delta\mathcal{T}_2\,\psi_{\Phi\bar{\Phi}}\,,\\
&\left[\bar{\mathcal{T}}_2(u)-\delta\bar{\mathcal{T}}_2-(u-u_1+i\xi_\Phi)\,(u-u_2-i\xi_\Phi)\right]\delta\psi_{\Phi\bar{\Phi}} = -\,\delta\bar{\mathcal{T}}_2\,\psi_{\Phi\bar{\Phi}}\,.
\end{aligned}
\end{equation}
The equations \eqref{FulleqRegular} for the non-singular part are self-contained and should be solved first. While these second-order differential equations generally have two independent solutions, we can dismiss one of them as non-physical by imposing a boundary condition that demands $\psi_{\Phi\bar{\Phi}}$ to be regular at $\sigma_1=\sigma_2$.

The term in the transfer matrix $\cT_2$ (\ref{Twf2}) that mixes different components of the wave function (\ref{wfansatz}) is proportional to $\theta_1\partial_{\theta_2}$.
Hence, $\cT_2$ does not mix the gluon component $\psi_{\Phi\bar{\Phi}}^{\bar FF}$ with other components.
After solving for $\psi_{\Phi\bar{\Phi}}^{\bar FF}$ we can move on to  $\psi_{\Phi\bar{\Phi}}^{\bar\psi\psi}$, with $\psi_{\Phi\bar{\Phi}}^{\bar FF}$ entering as a source. One could, potentially, proceed all the way to $\psi_{\Phi\bar{\Phi}}^{F\bar F}$ in this fashion.
Similarly, the term in the transfer matrix $\bar\cT_2$  that mixes different components of the wave function (\ref{wfansatz}) is proportional to $\theta_2\partial_{\theta_1}$. As a result, $\bar{\cT_2}$ does not mix the gluon component $\psi_{\Phi\bar{\Phi}}^{F\bar F}$ with other components and can be used
to find all the components in the opposite order. However, we found it simpler to stop halfway through at the scalar component $\psi_{\Phi\bar{\Phi}}^{\phi\bar{\phi}}$ in both cases, and obtain the full solution by merging the two pieces.
The resulting non-singular part of the wave function is given by
\begin{align}\la{2point35}
\psi_{\Phi\bar{\Phi}}^{\Upsilon\bar{\Upsilon}} =\frac{(-1)^{\delta_{\Upsilon\psi}}}{\mathbb{N}_{\Phi\bar{\Phi}}}\frac{8}{\lambda_\Phi(u_1)\,\lambda_{\bar{\Phi}}(u_2)}\Bigl[&e^{2iu_1\sigma_1+2iu_2\sigma_2+2|\xi_\Phi-\xi_{\Upsilon}|\left(\sigma_1-\sigma_2\right)}H^\Upsilon_{\bar{\Phi}}\Big(u_1,u_2\left|e^{2(\sigma_1-\sigma_2)}\Big)\right.
\\
&\!\!+e^{2iu_2\sigma_1+2iu_1\sigma_2+2|\xi_{\bar{\Phi}}-\xi_{\Upsilon}|\left(\sigma_1-\sigma_2\right)}\,H^{\Upsilon}_{\Phi}\left(-u_1,-u_2\left|e^{2\left(\sigma_1-\sigma_2\right)}\right)\right.\Bigr],\nn
\end{align}
where
\begin{align}
H^{F}_{\Phi}(u_1,u_2|r) &=\Gamma\left(\tfrac{1}{2}+\xi_\Phi+iu_1\right)\Gamma\left(\tfrac{1}{2}+\xi_\Phi-iu_2\right)\Gamma\left(iu_2-iu_1-2\xi_\Phi\right)\nn\\
&\phaneq\times{}_2F_1\left(\left.\tfrac{1}{2}+\xi_\Phi+iu_1,\tfrac{1}{2}+\xi_\Phi-iu_2,1+2\xi_\Phi+iu_1-iu_2\right|r\right) ,\nn\\
H^{\psi}_{\Phi}(u_1,u_2|r) &=\Gamma\left(\tfrac{3}{2}+\xi_\Phi+iu_1\right)\Gamma\left(\tfrac{1}{2}+\xi_\Phi-iu_2\right)\Gamma\left(iu_2-iu_1-2\xi_\Phi\right)\nn\\
&\phaneq\times{}_2F_1\left(\left.\tfrac{3}{2}+\xi_\Phi+iu_1,\tfrac{1}{2}+\xi_\Phi-iu_2,1+2\xi_\Phi+iu_1-iu_2\right|r\right)\nn\\
&\phaneq+\frac{1}{2}\left(1-i(1-\xi_\Phi)u_1-i(1+\xi_\Phi)u_2\right)H^{F}_{\Phi}(u_1,u_2|r)\, ,\nn\\
H^{\phi}_{\Phi}(u_1,u_2|r)
&=\Gamma\left(\tfrac{3}{2}+\xi_\Phi+iu_1\right)\Gamma\left(\tfrac{3}{2}+\xi_\Phi-iu_2\right)\Gamma\left(iu_2-iu_1-2\xi_\Phi\right)\nn\\
&\phaneq\times{}_2F_1\left(\left.\tfrac{3}{2}+\xi_\Phi+iu_1,\tfrac{3}{2}+\xi_\Phi-iu_2,1+2\xi_\Phi+iu_1-iu_2\right|r\right)\nn\\
&\phaneq+\frac{1}{12}\big(1 - 2\,(u_1^2+u_2^2+4u_1u_2)- 6i\xi_\Phi(u_1-u_2)+2\xi_\Phi^2\left((u_1-u_2)^2-2\right)\big) \nn\\
&\phaneq\quad \times H^{F}_{\Phi}(u_1,u_2|r)+ i\xi_\Phi\left(u_1-u_2\right)H^{\psi}_{\Phi}(u_1,u_2|r)\, ,
\end{align}
and
\begin{align}
H^{\bar{\Upsilon}}_{\Phi}(u_1,u_2|r) = H^{\Upsilon}_{\bar{\Phi}}(u_1,u_2|r)\,.
\end{align}
Additional normalization factors are given by
\begin{equation}
\begin{aligned}
\mathbb{N}_{\bar{F}F} &= \mathbb{N}_{F\bar{F}} = \Gamma\left(-\,\tfrac{1}{2}+iu_1\right)\Gamma\left(-\,\tfrac{1}{2}-iu_2\right)\Gamma\left(3-iu_1+iu_2\right)\, ,\\
\mathbb{N}_{\bar{\psi}\psi} &= -\,\mathbb{N}_{\psi\bar{\psi}} = \frac{i}{4}\,(u_1-u_2-2i)\,\Gamma\left(iu_1\right)\Gamma\left(-\,iu_2\right)\Gamma\left(1-iu_1+iu_2\right)\, ,\\
\mathbb{N}_{\phi\bar{\phi}} &= -\,\frac{1}{6}\,(u_1-u_2-2i)\,(u_1-u_2-i)\,\Gamma\left(\tfrac{1}{2}+iu_1\right)\Gamma\left(\tfrac{1}{2}-iu_2\right)\Gamma\left(-iu_1+iu_2\right)\, .
 \end{aligned}
\end{equation}
Lastly, the quantity $\lambda_{\Phi}(u)$ is given by
\begin{equation}\label{lambdadef}
\lambda_{\Phi}(u) = \frac{\pi}{{\rm cosh}(\pi(u+i\xi_\Phi))}\, .
\end{equation}

With the solutions for the regular part of the wave function constructed, we can solve the remaining  equations \eqref{FulleqSingular} that determine the singular part $\delta\psi_{\Phi\bar{\Phi}}$ of the corresponding wave function. We find
\beqa
\delta\psi_{\Phi\bar{\Phi}}= \frac{16}{\mathbb{N}_{\Phi\bar{\Phi}}}\,e^{i\left(u_1+u_2\right)\left(\sigma_1+\sigma_2\right)}\!&\Big[&\!\delta(\sigma_1-\sigma_2)\left(\theta_1^{[1}\theta_2^{\vphantom{[}2}\theta_2^{\vphantom{[}3}\theta_2^{4]} -  \theta_1^{[1}\theta_1^{\vphantom{[}2}\theta_1^{\vphantom{[}3}\theta_2^{4]}\right) \\
&&\!\!+\,\tfrac{3}{2}\left(\left(u_1 u_2+\xi_\Phi^2-\tfrac{3}{4}\right)\delta(\sigma_1-\sigma_2) - \tfrac{1}{2}\,\delta'(\sigma_1-\sigma_2)\right)\theta_1^{[1}\theta_1^{\vphantom{[}2}\theta_2^{\vphantom{[}3}\theta_2^{4]}\Bigr].\nn
\eeqa

The above wave functions are implemented in the ancillary file \texttt{Born level form factor OPE.nb} attached to this publication. 

\subsection*{Testing the solution}
We have preformed two types of tests of the solution. Firstly, from the asymptotic limit $\sigma_1\ll \sigma_2$, one can read off the S-matrices in the singlet sector as
\begin{align}
&\lim_{(\sigma_2-\sigma_1)\to\infty}\int d^8\theta\,\Psi_{\Phi\bar\Phi}(\boldsymbol{\sigma},\boldsymbol{\theta}|\textbf{u})|\boldsymbol{\sigma},\boldsymbol{\theta}\rangle\\
&=\frac{8}{\lambda_\Phi(u_1)\,\lambda_{\bar{\Phi}}(u_2)}\[e^{2i(u_1\sigma_1+u_2\sigma_2)}|\Phi_1,\bar\Phi_2\> + S_{\Phi\bar\Phi}(u_1,u_2)\,e^{2i(u_2\sigma_1+u_1\sigma_2)}|\bar\Phi_1,\Phi_2\>\]\, ,\nn
\end{align}
where the state $|\Phi_1,\bar\Phi_2\>$ is a shorthand notation for
\beq\la{tpstate}
|\Phi_1,\bar\Phi_2\>\equiv \left(x'(\sigma_1)x'(\sigma_2)\right)^{\xi_\Phi+\frac{1}{2}}\  \star\,\Phi(x(\sigma_1))\,\star\,\bar\Phi(x(\sigma_2))\,\star\, .
\eeq
We find 
\begin{align}\label{ssm}
S_{F\bar{F}}(u_1,u_2) &= \lim_{r\to 0}\frac{H^F_{\bar{F}}(-u_1,-u_2|r)}{H^F_{\bar{F}}(u_1,u_2|r)} = \frac{u_1-u_2-i}{u_1-u_2+i}\,\mathcal{S}_\frac{3}{2}(u_1,u_2)\, ,\nonumber\\
S_{\psi\bar{\psi}}(u_1,u_2) &= -\,\lim_{r\to 0}\frac{H^\psi_{\bar{\psi}}(-u_1,-u_2|r)}{H^\psi_{\bar{\psi}}(u_1,u_2|r)} = -\,\frac{u_1-u_2+2i}{u_1-u_2-2i}\,\mathcal{S}_1(u_1,u_2)\, ,\\
S_{\phi\bar{\phi}}(u_1,u_2) &= \lim_{r\to 0}\frac{H^\phi_{\phi}(-u_1,-u_2|r)}{H^\phi_{\phi}(u_1,u_2|r)} = \frac{(u_1-u_2+i)\,(u_1-u_2+2i)}{(u_1-u_2-i)\,(u_1-u_2-2i)}\,\mathcal{S}_\frac{1}{2}(u_1,u_2)\, ,\nonumber
\end{align}
where
\begin{align}
\mathcal{S}_s(u_1,u_2) = \frac{\Gamma\left(s-iu_1\right)\Gamma\left(s+iu_2\right)\Gamma\left(iu_1-iu_2\right)}{\Gamma\left(s+iu_1\right)\Gamma\left(s-iu_2\right)\Gamma\left(iu_2-iu_1\right)}\, ,
\end{align}
is the $S$-matrix of two identical excitations of conformal spin $s$. These S-matrices are in perfect agreement with the known GKP S-matrices derived in \cite{Basso:2010in}.

The second test we have preformed is to verify that these wave functions indeed lead to the correct factorized square and pentagon transitions \cite{Basso:2014koa}. This test is presented in appendices \ref{SWFappendix} and \ref{PTappendix}. We note here that our normalization of the wave functions, with the additional factor of $\frac{8}{\lambda_\Phi(u_1)\,\lambda_{\bar{\Phi}}(u_2)}$, is chosen for it to be consistent with the one of the square and pentagon transitions in \cite{Basso:2014koa}.

\section{The Form Factor Transitions}\la{FFsec}

With the explicit two-particle singlet states at Born level at hand, we can now compute the Born-level FF transitions of these states, i.e.\ the amplitude for such a state to be absorbed by the two-sided wrapped Wilson loop, see figure \ref{FFtransdefinition}. 

\begin{figure}[t]
\centering
\includegraphics[width=10cm]{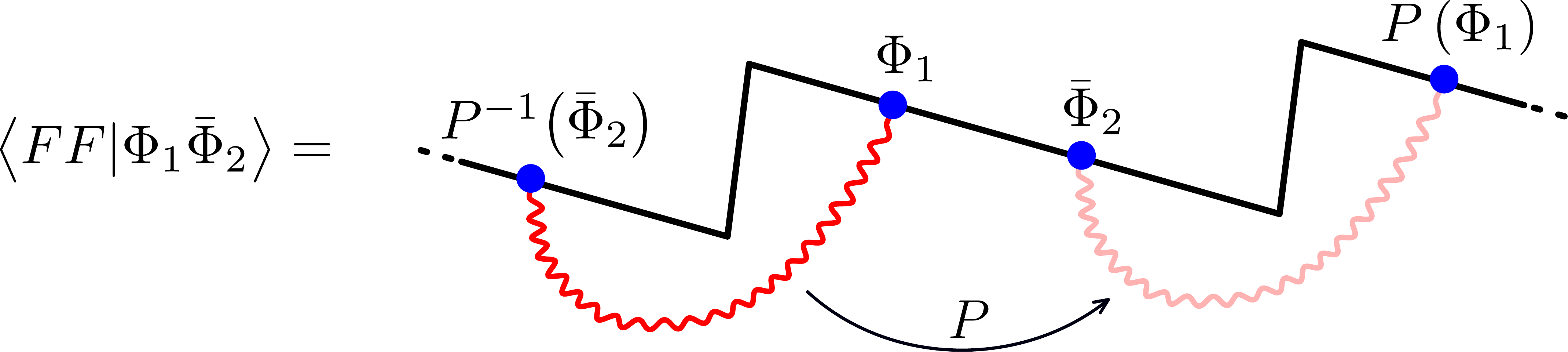}
\caption{At Born level, the FF transition for a pair of conjugate fields, ${\Phi}_1=\Phi(x(\sigma_1))$ and $\bar\Phi_2=\bar\Phi(x(\sigma_2))$ is given by the free propagator between $\bar\Phi(x(\sigma_2))$ and $\Phi\(P(x(\sigma_1))\)$.}
\label{tree4}
\end{figure}

Recall that each of these states is a certain superposition of two conjugate fields inserted on the edge of the Wilson loop (\ref{tpstate}). At Born level, we are instructed to contract them by a free propagator. This planar propagator emerges as the Wick contraction between one field on an edge of the wrapped two-sided polygon and the other field on a periodic image of the same edge, see figure \ref{tree4}. 
The result is
\begin{align}
\<FF|\Phi_1,\bar\Phi_2\>\equiv&\left(\d_{\sigma_1} x(\sigma_1)\d_{\sigma_2} P(x(\sigma_2))\right)^{\xi_\Phi+\frac{1}{2}}\,
\contraction{}{\Phi}{(P(x(\sigma_1)))\,}{\bar\Phi}\Phi(P(x(\sigma_1)))\,\bar\Phi(x(\sigma_2))\\
=&\(e^{\sigma_1+\sigma_2}+e^{-\sigma_1-\sigma_2}+2\,e^{\sigma_1-\sigma_2}\)^{-2\left(\xi_\Phi+\frac{1}{2}\right)}\, .\nn
\end{align}
If instead we use the superfield convention (\ref{statedef}), the same tree-level contraction takes the form
\begin{align}\label{1pptr}
\langle{\rm FF}|\sigma_1,\sigma_2;\theta_1,\theta_2\rangle =\,\,& \frac{\theta_1^{\vphantom{[}1}\theta_1^{\vphantom{[}2}\theta_1^{\vphantom{[}3}\theta_1^{\vphantom{[}4} +\theta_2^{\vphantom{[}1}\theta_2^{\vphantom{[}2}\theta_2^{\vphantom{[}3}\theta_2^{\vphantom{[}4}}{\left(e^{\sigma_1+\sigma_2}+e^{-\sigma_1-\sigma_2}+2\,e^{\sigma_1-\sigma_2}\right)^3} - 2\,\frac{\theta_1^{[1}\theta_1^{\vphantom{[}2}\theta_1^{\vphantom{[}3}\theta_2^{4]} - \theta_1^{[1}\theta_2^{\vphantom{[}2}\theta_2^{\vphantom{[}3}\theta_2^{4]}}{\left(e^{\sigma_1+\sigma_2}+e^{-\sigma_1-\sigma_2}+2\,e^{\sigma_1-\sigma_2}\right)^2}\nonumber\\
&+ 3\,\frac{\theta_1^{[1}\theta_1^{\vphantom{[}2}\theta_2^{\vphantom{[}3}\theta_2^{4]}}{\left(e^{\sigma_1+\sigma_2}+e^{-\sigma_1-\sigma_2}+2\,e^{\sigma_1-\sigma_2}\right)}\, .
\end{align}

The form factor transitions for the singlet states are obtained by convoluting (\ref{1pptr}) with the wave functions as
\begin{align}\label{FFxi1}
\<FF|&\Psi_{\Phi\bar\Phi}(u_1,u_2)\>
= \int d\sigma_1d\sigma_2d^4\theta_1d^4\theta_2\,\langle{\rm FF}|\sigma_1,\sigma_2;\theta_1,\theta_2\rangle\,\Psi_{\Phi\bar{\Phi}}(\sigma_1,\sigma_2;\theta_1,\theta_2|u_1,u_2)\\
&= I^{\bar{F}F}_{\Phi\bar{\Phi}}(u_1,u_2) - 2\,I^{\bar{\psi}\psi}_{\Phi\bar{\Phi}}(u_1,u_2) + 3\,I^{\phi\bar{\phi}}_{\Phi\bar{\Phi}}(u_1,u_2) +2\,I^{\psi\bar{\psi}}_{\Phi\bar{\Phi}}(u_1,u_2) + I^{F\bar{F}}_{\Phi\bar{\Phi}}(u_1,u_2)\, ,\nonumber
\end{align}
where in the second line we have split the integral into the contributions of the five different components: 
\beq
I^{\bar{a}a}_{\Phi\bar{\Phi}}(u_1,u_2)\equiv\int\limits_{\sigma_1<\sigma_2} d\sigma_1d\sigma_2\,\frac{\Psi_{\Phi\bar{\Phi}}^{\bar{a}a}(\sigma_1,\sigma_2|u_1,u_2)}{\left(e^{\sigma_1+\sigma_2}+e^{-\sigma_1-\sigma_2}+2\,e^{\sigma_1-\sigma_2}\right)^{2|\xi_a|+1}}\, .
\eeq

When both the regular and the singular part of the wave function are taken into account, one finds remarkably simple analytic expressions for these integrals:
\begin{align}\label{intsing}
I^{\bar{F}F}_{\Phi\bar{\Phi}}(u_1,u_2)&
= \frac{\pi}{6}\,\frac{(-1)^{2\xi_\Phi}}{\mathbb{N}_{\Phi\bar{\Phi}}}\,\frac{(u_1-u_2-2i{\xi_\Phi})\left((u_1-u_2-2i{\xi_\Phi})^2+1\right)}{{\rm sinh}(\pi(u_1-u_2))}\, ,\\
I^{\bar{\psi}\psi}_{\Phi\bar{\Phi}}(u_1,u_2)&=\frac{\pi}{3}\,\frac{(-1)^{2\xi_\Phi}}{\mathbb{N}_{\Phi\bar{\Phi}}}\,\frac{(u_1-u_2-2i{\xi_\Phi})\left((u_1-u_2)^3-i{\xi_\Phi}\,(u_1-u_2)+4(1-{\xi_\Phi}^2)\right)}{{\rm sinh}(\pi(u_1-u_2))}\, ,\nn\\
I^{\phi\bar{\phi}}_{\Phi\bar{\Phi}}(u_1,u_2)&= \frac{\pi}{3}\,\frac{(-1)^{2\xi_\Phi}}{\mathbb{N}_{\Phi\bar{\Phi}}}\,\frac{(u_1-u_2)\left((u_1-u_2)^2+1\right)\left((u_1-u_2)^2+4\right)}{\left((u_1-u_2)^2+4{\xi_\Phi}^2\right){\rm sinh}(\pi(u_1-u_2))}\, ,\nn
\end{align}
where $I^{F\bar{F}}_{\Phi\bar{\Phi}}$ is related to $I^{\bar FF}_{\Phi\bar{\Phi}}$ by flipping the sign of $\xi_\Phi$ and
$I^{\psi\bar\psi}_{\Phi\bar{\Phi}}$ is related to $I^{\bar{\psi}\psi}_{\Phi\bar{\Phi}}$ by a flip of the sign of $\xi_\Phi$ and an overall sign flip.

The integral combinations $I^{\bar{F}F}_{\Phi\bar{\Phi}}(u_1,u_2) + I^{F\bar{F}}_{\Phi\bar{\Phi}}(u_1,u_2)$ and $I^{\bar{\psi}\psi}_{\Phi\bar{\Phi}}(u_1,u_2) - I^{\psi\bar{\psi}}_{\Phi\bar{\Phi}}(u_1,u_2)$ that enter (\ref{FFxi1}) are finite at $u_1 = u_2$. However, in each of these combinations the two terms individually have a simple pole at this point, which cancel upon addition. These cases require an $i\epsilon$ prescription that follows from our choice of tessellation, see the discussion in section 6 of \cite{Basso:2014koa}. 
As a result, the poles do not cancel completely, but leave behind a delta-function contribution. More specifically,
\begin{align}
&I^{\bar{F}F}_{F\bar{F}}(u_1-i\epsilon,u_2+i\epsilon) + I^{F\bar{F}}_{F\bar{F}}(u_1+i\epsilon,u_2-i\epsilon) = -\,2\left(u_1^2+\frac{1}{4}\right){\rm cosh}(\pi u_1)\,\delta\left(u_1-u_2\right)\nonumber\\
&+ \frac{i}{3}\,\frac{(u_1-u_2)^2-11}{u_1-u_2+i}\,\frac{\Gamma\left(iu_1-iu_2+1\right)}{\Gamma\left(-\,\frac{1}{2}+iu_1\right)\Gamma\left(-\,\frac{1}{2}-iu_2\right)} + \mathcal{O}(\epsilon)\, ,
\end{align}
and
\begin{align}
&I^{\bar{\psi}\psi}_{\psi\bar{\psi}}(u_1-i\epsilon,u_2+i\epsilon) - I^{\psi\bar{\psi}}_{\psi\bar{\psi}}(u_1+i\epsilon,u_2-i\epsilon) = u_1\,{\rm sinh}(\pi u_1)\,\delta\left(u_1-u_2\right)\nonumber\\
&- \frac{i}{3}\,\frac{(u_1-u_2)^2+\frac{5}{2}}{u_1-u_2-2i}\,\frac{\Gamma\left(iu_1-iu_2+1\right)}{\Gamma\left(iu_1\right)\Gamma\left(-\,iu_2\right)} + \mathcal{O}(\epsilon)\, .
\end{align}

After adding up the contributions of all the components, one finds the following simple expressions for the Born-level form factor transitions:
\begin{align}
&F_{\phi\bar{\phi}}(u_1,u_2) = -\,6\times\frac{4}{g^2\,(u_1-u_2-2i)\,(u_1-u_2-i)}\,\frac{\Gamma\left(iu_1-iu_2\right)}{\Gamma\left(\frac{1}{2}+iu_1\right)\Gamma\left(\frac{1}{2}-iu_2\right)}\, ,\nonumber\\
&F_{\psi\bar{\psi}}(u_1,u_2) = 4\times\frac{2}{g^2}\,u_1\,{\rm sinh}(\pi u_1)\,\delta(u_1-u_2)\, ,\label{FFBorn}\\
&F_{F\bar{F}}(u_1,u_2) = -\,1\times\frac{2}{g^2}\left(u_1^2+\frac{1}{4}\right){\rm cosh}(\pi u_1)\,\delta(u_1-u_2)\, ,\nonumber
\end{align}
where $F_{\Phi\bar\Phi}\equiv\<FF|\Psi_{\Phi\bar\Phi}\>/g^2$.
Note that the prefactors $6$, $4$ and $1$ count the numbers of real scalars, complex fermions and complex gauge fields, respectively. We have included them in the form factor transition for convenience.\footnote{Alternatively, they can be accounted for by summing over the matrix indices of the form factor transitions \cite{Sever:2020jjx}, which we have suppressed throughout this paper.} 
Remarkably, when all the components are added together the parts of the gluon and fermion transitions that are regular at $u_1=u_2$ completely cancel, leaving only the $\delta$-function contributions times the corresponding inverse measure:
\begin{equation}\la{eq:FFBornintermsofmeasures}
\begin{aligned}
 &F_{\psi\bar{\psi}}(u_1,u_2) = 4\times\frac{2\pi}{\mu_{\psi}(u_1)}\,\delta(u_1-u_2)\, ,\\
&F_{F\bar{F}}(u_1,u_2) = 1\times\frac{2\pi}{\mu_F(u_1)}\,\delta(u_1-u_2)\, .
\end{aligned}
\end{equation}
The existence of this $\delta$-function contribution is required by the square-limit axiom put forward in \cite{Sever:2020jjx}. Here, we see that the full transition for the two gluons and two fermions states is given solely by this contribution.
In fact, we will argue in \cite{Toappear2} that relation \eqref{eq:FFBornintermsofmeasures} essentially remains true also at finite coupling.

\section{Matching with Data}\la{data}

With the FF transitions at hand, we can now produce OPE predictions for the FF itself. 

We consider the conformal invariant ratio ${\cal W}_3$ defined in figure \ref{ratiofig}. In the large-$\tau$ collinear limit, the leading contribution to the OPE expansion of this object (\ref{W3}) at weak coupling comes from the three two-particle singlet states we have studied above. Their contribution is given by
\beq\la{W3-2pt}
\cW_3
=\sum\limits_{\Phi\in\{F,\psi,\phi\}}\!\!\mathcal{N}_\Phi\int\frac{du\,dv}{(2\pi)^2}\,e^{ip_{\Phi\bar\Phi}\sigma-E_{\Phi\bar\Phi}\tau}P_{\Phi\bar{\Phi}}(0|u,v)\,\mu_\Phi(u)\,\mu_{\bar{\Phi}}(v)\,F_{\Phi\bar{\Phi}}(u,v)+\mathcal{O}(e^{-4\tau})\, ,
\eeq
where  the symmetry factor $\mathcal{N}_\Phi$ is equal to $1/2$ when $\Phi = \phi$ and $1$ otherwise. 

We now expand the elements entering (\ref{W3-2pt}) at leading order in perturbation theory. Firstly, the tree-level energy is equal to the twist, $E_{\Phi\bar\Phi}=2+\cO(g^2)$. This means that at $\ell$-loop order, the $\tau$-dependence of the leading OPE contribution takes the form $\cW_3^{(\ell)}=e^{-2\tau}P_{\ell-1}(\tau)+\cO(e^{-4\tau}\tau
^{\ell-1})$ where $P_{\ell-1}$ is a polynomial of degree $\ell-1$.
Secondly, the momentum is given by 
$p_{\Phi\bar\Phi}=2u+2v+\cO(g^2)$. Thirdly, the measures for the particles read~\cite{Basso:2014koa}
\begin{equation}\label{measures}
\begin{aligned}
\mu_\phi(u) &=\mu_{\bar\phi}(u) = \frac{\pi g^2}{{\rm cosh}(\pi u)}+\cO(g^4)\, , \\
\mu_\psi(u)&=\mu_{\bar\psi}(u)= \frac{\pi g^2}{u\,{\rm sinh}(\pi u)}+\cO(g^4)\, ,\\
\mu_F(u)&=\mu_{\bar F}(u)= -\,\frac{\pi g^2}{\left(u^2+\frac{1}{4}\right){\rm cosh}(\pi u)}+\cO(g^4)\, .
\end{aligned}
\end{equation}
And lastly, $P_{\Phi\bar{\Phi}}(0|u_1,u_2)$, the pentagon transitions responsible for creating the two-particle singlet states from the OPE vacuum, are given by \cite{Basso:2014koa}\footnote{These objects can be divided into a $g$-dependent kinematical part and a $g$-independent matrix part. Since we are only concerned with their leading order value in this paper, this split is not relevant to us.}
\begin{equation}\label{creation-amplitudes}
\begin{aligned}
&P_{\phi\bar{\phi}}(0|u_1,u_2) = \frac{1}{(u_1-u_2+i)\,(u_1-u_2+2i)}\times\frac{\Gamma\left(\frac{1}{2}+iu_1\right)\Gamma\left(\frac{1}{2}-iu_2\right)}{\Gamma\left(iu_1-iu_2\right)}+\cO(g^2)\, ,\\
&P_{\psi\bar{\psi}}(0|u_1,u_2) = \frac{1}{u_1-u_2+2i}\times\frac{\Gamma\left(1+iu_1\right)\Gamma\left(1-iu_2\right)}{\Gamma\left(1+iu_1-iu_2\right)}+\cO(g^2)\, ,\\
&P_{F\bar{F}}(0|u_1,u_2) = \frac{\Gamma\left(\frac{3}{2}+iu_1\right)\Gamma\left(\frac{3}{2}-iu_2\right)}{\Gamma\left(2+iu_1-iu_2\right)}+\cO(g^2)\, .
\end{aligned}
\end{equation}

Inserting all the factors into (\ref{W3-2pt}), together with the FF transitions derived above (\ref{FFBorn}), we  can preforming the integrations over $u$ and $v$ in a series expansion in $e^{\pm 2\sigma}$ via residues,%
\footnote{Note that for fermions, there is a pole on the real integration axis, which has to be shifted as prescribed in \cite{Basso:2014koa}.}
and resum to arrive at the first OPE prediction
\beq\la{OPEpredict}
\mathcal{W}_{3}^{(1)}=2e^{-2\tau}\left(1 - 2\sigma\,e^{-2\sigma} - 4\cosh^2(\sigma)\log\left(1+e^{-2\sigma}\right)\right)
+O(e^{-4\tau})\, ,
\eeq
where $\mathcal{W}_{3}=1+\sum_{\ell=1}^\infty g^{2\ell}\,\mathcal{W}_{3}^{(\ell)}$.

In order to test this prediction, we extract $\mathcal{W}_{3}^{(1)}$ from previously computed form factors~\cite{Brandhuber:2010ad}. At one-loop order, one finds
\begin{align}\label{RRU1}
\mathcal{W}_{3}^{(1)} =2\[{\rm Li}_2(-e^{-2\tau} - e^{2\sigma})+ {\rm Li}_2(-e^{-2\tau} - e^{-2\sigma}(1 + e^{-2\tau})^2)-{\rm Li}_2(-e^{-2\tau})+2\sigma^2 +\frac{\pi^2}{6}\]\,.
\end{align}
Expanding (\ref{RRU1}) to leading order in $e^{-2\tau}$, we find a perfect agreement with the OPE prediction (\ref{OPEpredict}). Higher powers of $e^{-\tau}$ in the expansion of (\ref{RRU1}) come from OPE states with more than two particles. We note that these come in even powers of $e^{-\tau}$ only. This selection rule has a simple all-loop OPE explanation that is based on the type of GKP excitations and the singlet constraint, see \cite{Sever:2020jjx}. 

Next, using the FF transitions at Born level only, we can already make a prediction for a piece of the FF at any loop order.  
Namely, we can predict the leading $\ell$'th power of $\tau$ at $(\ell+1)$-loop order,
\beq
\mathcal{W}^{(\ell+1)}_{3}(\tau,\sigma)=\tau^\ell e^{-2\tau}\times\mathcal{W}^{(\ell+1)}_{3,\,\tau^{\ell}e^{-2\tau}}(\sigma)+\cO(\tau^{\ell-1}e^{-2\tau})\, .
\eeq
The function $\mathcal{W}^{(\ell+1)}_{3,\,\tau^{\ell}e^{-2\tau}}(\sigma)$ is obtained by pulling down $\ell$ powers of the one-loop correction to the energy (\ref{onelE})
from the exponent $e^{-\tau(E_{s}(u_1)+E_{s}(u_2))}$. This leads to the following integral:
\begin{align}\la{leadingtau}
\mathcal{W}^{(\ell+1)}_{3,\,\tau^{\ell}e^{-2\tau}}(\sigma) &= \sum\limits_{\Phi\in\{F,\psi,\phi\}}\mathcal{N}_\Phi\int\frac{du\,dv}{(2\pi)^2}\,e^{2i\sigma(u+v)}\frac{(-1)^{\ell}}{ \ell!}\Big(E^{(1)}_{\xi_\Phi+\frac{1}{2}}(u) + E^{(1)}_{\xi_{\Phi}+\frac{1}{2}}(v)\Big)^{\ell}\\
&\qquad\qquad\qquad\qquad\qquad\times \left[g^{-2}P_{\Phi\bar{\Phi}}(0|u,v)\,\mu_\Phi(u)\,\mu_{\bar{\Phi}}(v)\,F_{\Phi\bar{\Phi}}(u,v)\right]_{g=0}\,,\nn
\end{align}

We find that the above loop corrections at any loop order can be written in terms of harmonic polylogarithms (HPLs) \cite{Remiddi:1999ew}:
\begin{equation}
 H_{a_1,\dots,a_n}\equiv H(a_1,\dots,a_n;-e^{-2\sigma})\,.
\end{equation}
HPLs can be conveniently manipulated using e.g.\ the \textsc{Mathematica} package \texttt{HPL} \cite{Maitre:2005uu}.

Using \eqref{RtoW}, we have obtained closed expressions for $\mathcal{W}^{(\ell)}_{3,\,\tau^{\ell-1} e^{-2\tau}}$ up to seven-loop order, which we provide in the auxiliary file \texttt{OPE-predictions.txt} attached to this paper.
For space reasons, we only give the results up to four-loop order in the main text:\footnote{At two- and  three-loop order, these expressions can also be written in terms of classical polylogarithms~\cite{Sever:2020jjx}.}
\begingroup
\allowdisplaybreaks
\begin{align}\label{eq:two-loop-OPE-prediction}
\mathcal{W}^{(2)}_{3,\,\tau^{1} e^{-2\tau}} &=
-\,64 \cosh ^2(\sigma) \left(H_{1,1} + \left(\frac{1}{2}+\sigma\right)H_1\right) - 32\,e^{\sigma} \sigma  \cosh (\sigma) - 8\, ,\\
\mathcal{W}^{(3)}_{3,\,\tau^{2} e^{-2\tau}}&= 64 \cosh ^2(\sigma) \Bigg(\left(3\sigma^2+4\sigma+2+\frac{\pi^2}{12}\right) H_1 + 8 \left(\frac{1}{2}+\sigma\right)H_{1,1}\nn\\*
&\phaneq+8 H_{1,1,1}+\frac{1}{2}H_{0,0,1}\Bigg) +128\,e^{\sigma }\sigma\cosh(\sigma ) + 48 \sigma^2 + \frac{4 \pi^2}{3}+24\, ,\\
\mathcal{W}^{(4)}_{3,\,\tau^{3} e^{-2\tau}} &= -\,64\cosh^2(\sigma)\Bigg[\left(\frac{40}{9}\sigma^3+\frac{19}{3} - 2\zeta_3\right)H_1 + \left(12\sigma^2 + \frac{\pi^2}{3}\right)\left(H_1+2H_{1,1}\right)\nn\\*
&\phaneq + 4\sigma\bigg(8H_{1,1} -\frac{1}{3} H_{0,0,1} + H_{0,1,1} + H_{1,0,1} + 16H_{1,1,1} + \left(\frac{7}{2}+\frac{\pi^2}{9}\right)H_1\bigg)\nn\\&\phaneq - H_{0,1} + 16H_{1,1} - 2H_{0,0,1} + 32H_{1,1,1} - \frac{20}{3}H_{0,0,1,1} - \frac{10}{3}H_{0,1,0,1} - \frac{10}{3}H_{1,0,0,1}\nn\\*&\phaneq
+ 64 H_{1,1,1,1}\Bigg] -64\,e^{\sigma}\cosh(\sigma)\left(H_{0,1} + 2 H_1 +3\sigma^2 +\frac{19}{3}\sigma +\frac{\pi^2}{12}\right)-\frac{640}{9} \sigma ^3\nn\\*
&\phaneq-96 \sigma ^2 - 64\sigma\left(H_{0,1}+\frac{\pi^2}{9}\right)-\frac{32}{3}H_{0,0,1}+32 \zeta_3-\frac{8\pi^2}{3}-\frac{160}{3}\, .
\end{align}
\endgroup

The two-loop data available for the three-point form factor reminder function ${\cal R}_3$ \cite{Brandhuber:2012vm} allows us to test the OPE prediction for the $\tau e^{-2\tau}$ term in ${\cal W}_3^{(2)}$. To do so, in general one first has to translate between these two finite dual conformally invariant functions, ${\cal R}$ and ${\cal W}$. The relation between them reads
\beq\la{RtoW}
{\cal W}_n=\exp\left[\frac{\Gamma_{\text{cusp}}}{4}\,\mathcal{W}^{(1)}_{n}\right]\times \mathcal{R}_n\, ,
\eeq
where $\Gamma_{\text{cusp}}=4g^2+\dots$ is the cusp anomalous dimension.
For the $\mathcal{W}^{(\ell)}_{3,\,\tau^{\ell-1} e^{-2\tau}}$ terms, this translation is trivial though:
\begin{equation}
 \mathcal{W}^{(\ell)}_{3,\,\tau^{\ell-1} e^{-2\tau}}=\mathcal{R}^{(\ell)}_{3,\,\tau^{\ell-1} e^{-2\tau}} \quad\text{for}\quad \ell\geq2\,.
\end{equation}
The two-loop result \eqref{eq:two-loop-OPE-prediction} perfectly agrees with the corresponding term in the three-point form factor reminder function ${\cal R}_3$  \cite{Brandhuber:2012vm}.

Our three-, four-, five-, six-, and seven-loop predictions have already provided valuable input for the perturbative form factor bootstrap up to seven-loop order \cite{Dixon:2020bbt,PerturbativeBootstrap2}.\footnote{In particular, the terms up to five-loop order in our auxiliary files are identical to those given for the remainder function up to five-loop order in the auxiliary files of \cite{Dixon:2020bbt}, where also some details on a more efficient resummation of the series expansion in $e^{\pm\sigma}$ in terms of HPLs are described.
}

\section{Discussion and outlook}\la{Conclusions}

In this paper, we have calculated the leading OPE contribution to the Born-level three-gluon form factor of the chiral half of the stress tensor supermultiplet in planar ${\cal N}=4$ SYM theory. To achieve this, we first constructed the previously unknown two-particle singlet states of the GKP flux tube. We then used them to evaluate the corresponding OPE form factor transitions, providing details that we deferred in \cite{Sever:2020jjx}. In addition to the leading FFOPE contribution, these transitions also provide us with the terms of order $e^{-2\tau}\tau^{\ell-1}$ at $\ell$ loops; we attach these predictions up to seven-loop order in the auxiliary file  \texttt{OPE-predictions.txt}. These predictions have already been used, combined with other data, for a perturbative bootstrap of the three-point form factor at any value of $\tau$ and $\sigma$ up to seven-loop order \cite{Dixon:2020bbt,PerturbativeBootstrap2}.
In \cite{Toappear2}, we will use a bootstrap procedure to determine the two-particle form factor transitions at finite coupling, thus lifting the Born-level results of this paper to any value of the coupling.

The leading OPE contribution that we have considered is of order $e^{-2\tau}$, resulting from states of tree-level energy or twist 2.
As we move to higher twist, and correspondingly higher powers of $e^{-2\tau}$ of the Born-level FF, we encounter 
two types of two-particle singlet states. The first type consists of a gluon bound state, $D_z^nF_{z-}\equiv F_{n+1}$, and its complex conjugate, $\bar F_{n+1}$, which are considered to be fundamental excitations. The other type consists of two effective (or one effective and one fundamental) excitations \cite{Basso:2010in}. GKP states with more than two particles only start contributing at higher loop orders. 
Below, we give conjectures for the Born-level form factor transitions of the two aforementioned state types that contribute at order $e^{-4\tau}$. 

Let us first consider a singlet state made of two gluon bound states $F_a\bar F_a$. The gluon form factor  transition, constructed in (\ref{eq:FFBornintermsofmeasures}), corresponds to $a=1$. A natural generalization of it to higher $a$ is obtained by simply replacing the measure $\mu_F\equiv \mu_{F_{1}}$ with the bound state measure $\mu_{F_{a}}$ given in \cite{Basso:2014nra}: 
\begin{equation}\la{boundstateFF}
 F_{F_a\bar{F}_a}(u_1,u_2) = \frac{2\pi}{\mu_{F_a}(u_1)}\,\delta(u_1-u_2)\, .
\end{equation}
We expect this generalization to hold at higher loops orders as well.

Next, we consider a singlet state made of two effective (or one effective and one fundamental) excitations. The finite-coupling counterparts of the effective excitations are not, in fact, single-particle states. Instead, they are multi-particle states that consist of a fundamental excitation with finite momentum and a set of fermions with momenta smaller than the 't Hooft coupling, called {\it small fermions}. 
At weak coupling, the small fermions attach themselves to finite-momentum excitation and act as supersymmetry generators \cite{Alday:2007mf}. This mechanism results in a sea of effective single-particle excitations at weak coupling, see the discussions in \cite{Basso:2014koa,Cordova:2016woh}. Hence, one way of obtaining the FF transitions for these effective excitations is to start with the finite-coupling expressions for the multi-particle FF transitions with small fermions and take the weak-coupling limit. Finite-coupling multi-particle FF transitions are subject to a set of axioms put forward in \cite{Sever:2020jjx}. Constructing a finite-coupling solution to these axioms is beyond the scope of this paper, though.
Instead, here we will follow a (conjectural) shortcut which turns out to work much better than expected. 

Among the set of finite-coupling axioms, only one -- the so-called mirror axiom -- does not have a weak-coupling counterpart. It is not hard to check that the mirror axiom is also the only one not satisfied by a simple factorized ansatz, presented below for four (fundamental) particles. This ansatz for multi-particle FF transitions also happens to capture the small fermion attachment mechanism. 
It therefore leads to Born-level transitions of two effective excitations that satisfy all axioms that survive in the weak-coupling limit. To our surprise, we observe that the conjectures obtained in this way indeed match with the one-loop data at order $e^{-4\tau}$.
Moreover, with the known corrections to the energy of the flux-tube excitations, these conjectures correctly reproduce the terms of order $e^{-4\tau}\tau^{\ell-1}$ up to seven loops \cite{Brandhuber:2012vm,Dixon:2020bbt,PerturbativeBootstrap2}. 
Taking into account the all-loop two-particle transitions we derive in \cite{Toappear2}, we also find perfect agreement with the terms of order $e^{-4\tau}\tau^{\ell-2}$ and $e^{-4\tau}\tau^{\ell-3}$ up to seven loops \cite{Brandhuber:2012vm,Dixon:2020bbt,PerturbativeBootstrap2}.\footnote{Note that, in contrast to the $e^{-2\tau}$ terms, the $e^{-4\tau}$ terms have not been used as input for the perturbative bootstrap, so this is an independent test.} In particular, this means that we can reconstruct the full $e^{-4\tau}$ piece of the form factor up to three loops!\footnote{At four-loop order, we find that a four-scalar singlet state contributes, which is beyond the scope of this paper.} Hence, we have chosen to present this ansatz here.

There are three twist-4 states with fermions that can lead to a singlet state made of two effective (or one effective and one fundamental) excitations. They consist of two small fermions and either two gluons, two scalars or two large fermions. For either of these combinations, we order the excitations in such a way that the square limits \cite{Sever:2020jjx} 
can be taken directly, without needing to reorder the particles. For such an ordering, the factorized twist-4 ansatz takes the form
\begin{align}
\widetilde F_{F_1\psi_S\bar{\psi}_S\bar{F}_1}(u_1,v_1,v_2,u_2)&=F_{F_1\bar{F}_1}(u_1,u_2)F_{\psi_S\bar{\psi}_S}(v_1,v_2)\, ,\nn\\
\widetilde F_{\psi\psi_S\bar{\psi}_S\bar{\psi}}(u_1,v_1,v_2,u_2)&=F_{\psi\bar{\psi}}(u_1,u_2)F_{\psi_S\bar{\psi}_S}(v_1,v_2)\, ,\la{ansatz}\\
\widetilde F_{\psi_S\phi\bar{\phi}\bar{\psi}_S}(u_1,v_1,v_2,u_2)&=F_{\phi\bar{\phi}}(v_1,v_2)F_{\psi_S\bar{\psi}_S}(u_1,u_2)\, ,\nn
\end{align}
where $F_{\psi_S\bar{\psi}_S}$ is the form factor transition for two conjugate small fermions. Similarly to the gluon bound states (\ref{boundstateFF}), we expect it to take the form 
\begin{equation}
 F_{\psi_S\bar{\psi}_S}(u_1,u_2) = 4\times\frac{2\pi}{\mu_{\psi_S}(u_1)}\,\delta(u_1-u_2)\, ,
\end{equation}
where the measure $\mu_{\psi_S}$ can be found in \cite{Basso:2014koa}. In (\ref{ansatz}), the tilde in $\widetilde F$ indicates that the finite-coupling generalizations of these Born-level transitions are incompatible with the mirror axiom. To check this ansatz, the transitions \eqref{boundstateFF}--(\ref{ansatz}) have to be dressed by the pentagon creation amplitude for the same state, which includes the respective matrix part given in appendix~\ref{PTappendix}.

It would be interesting to check if analogous conjectures at higher orders in $e^{-2\tau}$ also agree with the Born-level form factor.
Another important direction is to understand why the ansatz ({\ref{ansatz}), that is not compatible with mirror axiom, matches the data so well?

While we have restricted ourselves to the three-point MHV form factor of the stress tensor supermultiplet in the present paper, the form factor transitions calculated here can also be used to provide FFOPE predictions for higher-point MHV form factors of the stress tensor supermultiplet, potentially also enabling a perturbative bootstrap in the higher-point cases.

It would be also interesting to calculate form factor transitions for other operators, for which the corresponding form factors have been studied in \cite{Engelund:2012re,Brandhuber:2014ica,Wilhelm:2014qua, Nandan:2014oga, Loebbert:2015ova, Brandhuber:2016fni, Loebbert:2016xkw, Caron-Huot:2016cwu,Banerjee:2016kri,Ahmed:2016vgl,Brandhuber:2018xzk,Lin:2020dyj}. We expect that the formalism developed in the present paper will also be useful in these cases.

\paragraph{Acknowledgments:} 
We are very grateful to B.~Basso for discussions. MW thanks L.~Dixon, {\"O}.~G{\"u}rdo{\u{g}}an and A.~McLeod for collaboration on \cite{Dixon:2020bbt,PerturbativeBootstrap2}. AT and MW are grateful to CERN for hospitality. AS is grateful to NBI for hospitality. AS was supported by the I-CORE Program of the Planning and Budgeting Committee, The Israel Science Foundation (grant number 1937/12) and by the Israel Science Foundation (grant number 1197/20). AT received funding from the European Research Council (ERC) under the European Unions Horizon 2020 research and innovation programme, Novel structures in scattering amplitudes (grant agreement No. 725110).
MW was supported in part by the ERC starting grant 757978 and the research grants 00015369 and 00025445 from Villum Fonden.

\appendix

\section{\texorpdfstring{$SL(2|4)$}{SL(2|4)} algebra}\label{SL24appendix}

In this appendix, we give the basic properties of the $SL(2|4)$ algebra. These include its commutation relations, Casimir operator and two of its simplest finite-dimensional representations, $\textbf{6}$ and $\bar{\textbf{6}}$, that are used for constructing the $R$-matrices. A detailed explanation of how the collinear supergroup emerges from the $PSU(2,2|4)$ generators of $\mathcal{N}=4$ SYM theory can be found in \cite{Belitsky:2004sc}.

The group generators transform in the adjoint representation of $SL(2|4)$, $\textbf{35}$. Under the reduction to $SL(2)\times SU(4)$ bosonic subgroup
\begin{align}
&\,\textbf{35}_{SL(2|4)} =\\
&\underbrace{\underbrace{\textbf{3}_{SL(2)}\otimes\textbf{1}_{SU(4)}}_{ L_{-},\,L_0,\,L_{+}} + \underbrace{\textbf{1}_{SL(2)}\otimes\textbf{15}_{SU(4)}}_{T_A{}^B} + \underbrace{\textbf{1}_{SL(2)}\otimes\textbf{1}_{SU(4)}}_B}_\text{Gra\ss{}mann-even} + \underbrace{\underbrace{\textbf{2}_{SL(2)}\otimes\textbf{4}_{SU(4)}}_{W_{-}^A,\,W_{+}^A} + \underbrace{\textbf{2}_{SL(2)}\otimes\bar{\textbf{4}}_{SU(4)}}_{V_{-,A},\,V_{+,A}}}_\text{Gra\ss{}mann-odd}.\nonumber
\end{align}
The generators of the $SL(2)$ subalgebra are chosen to satisfy the following commutation relations:\footnote{These commutation relations were chosen to mirror the $-1,0,1$ sector of the Virasoro algebra and lead to slightly unusual sign conventions.}
\begin{align}
\left[L_{-},\,L_{+}\right] = -\,2\,L_0\, , \quad \left[L_{0},\,L_{-}\right] = L_{-}\, , \quad \left[L_{0},\,L_{+}\right] = -\,L_{+}\, .
\end{align}
The commutators of the other generators with $L_i$ are determined by the $SL(2)$ representations they belong to,
\begin{align}
&\left[L_0,\,V_{\pm,A}\right] = \mp\tfrac{1}{2}\,V_{\pm,A}\, , && \left[L_0,\,W_{\pm}^A\right] = \mp\tfrac{1}{2}\,W_{\pm}^A\, ,\nonumber\\
&\left[L_\pm,\,V_{\mp,A}\right] = \pm\,V_{\pm,A}\, , && \left[L_\pm,\,W_{\mp}^A\right] = \pm\,W_{\pm}^A\, ,\nonumber\\
&\left[L_\pm,\,V_{\pm,A}\right] = 0\, , && \left[L_\pm,\,W_{\pm}^A\right] = 0\, ,\nonumber\\
&\left[L_i,\,T_{A}{}^B\right] = 0\, , && \left[L_i,\,B\right] = 0\, .
\end{align}
The auxiliary generator $B$ has the following commutation relations with the rest of the generators:
\begin{align}
\left[B,\,V_{\pm,A}\right] = \frac{1}{4}\,V_{\pm,A}\, ,\qquad \left[B,\,W_{\pm}^A\right] = -\,\frac{1}{4}\,W_{\pm}^A\, ,\qquad\left[B,\,T_{A}{}^B\right] = 0\, .
\end{align}
The action of the $SU(4)$ generators in the Gra\ss{}mann-odd part is \begin{align}
&\left[V_{\pm,A},\,T_B{}^C\right] = \delta_A^C\,V_{\pm,A} - \frac{1}{4}\,\delta_B^C\,V_{\pm,A}\, ,\qquad\left[W_{\pm}^A,\,T_B{}^C\right] = -\,\delta_A^B\,W_{\pm}^C + \frac{1}{4}\,\delta_B^C\,W_{\pm}^A\, ,
\end{align}
while the commutators of the $SU(4)$ generator with itself are
\begin{align}
&\left[T_A{}^B,\,T_C{}^D\right] = \delta_A^D\,T_C{}^B - \delta_C^B\,T_A{}^D\, .
\end{align}
Lastly, we complete the algebra by presenting the anticommutators of the Gra\ss{}mann-odd generators:
\begin{align}
&\{V_{\pm,A},\,W_{\pm}^B\} = \delta_{A}^B\,L_{\pm}\, ,\qquad \{V_{\pm,A},\,W_{\mp}^B\} = \mp\,T_{A}{}^B + \delta_{A}^B\left(L_0 \mp B\right)\, .
\end{align}

The Casimir operator of $SL(2|4)$ is essential for constructing the R-matrices. It is given by
\begin{align}\label{Cas}
\mathcal{C} =&\,\,L_0^2 - \frac{1}{2}L_{+}L_{-} - \frac{1}{2}L_{-}L_{+} + 2B^2 - \frac{1}{2}T_{A}{}^BT_{B}{}^A\nonumber\\
&\,\,- \frac{1}{2}\left(V_{-,A}\,W_{+}^A + W_{-}^A\,V_{+,A} - V_{+,A}\,W_{-}^A - W_{+}^A\,V_{-,A}\right)\, .
\end{align}

Another necessary piece in the construction of the R-matrix is the auxiliary space representation. The fundamental and anti-fundamental representations of $SL(2|4)$, $\textbf{6}$ and $\bar{\textbf{6}}$, reduce to the following representations of the $SL(2)\otimes SU(4)$ subgroup:
\begin{align}
\textbf{6}_{SL(2|4)} &= \textbf{2}_{SL(2)}\otimes\textbf{1}_{SU(4)} + \textbf{1}_{SL(2)}\otimes\textbf{4}_{SU(4)}\, ,\\
\bar{\textbf{6}}_{SL(2|4)} &= \textbf{2}_{SL(2)}\otimes\textbf{1}_{SU(4)} + \textbf{1}_{SL(2)}\otimes\bar{\textbf{4}}_{SU(4)}\, .\nonumber
\end{align}
These representations are conjugate to each other under the following automorphism of the $SL(2|4)$ algebra:
\begin{align}
&L_i \rightarrow L_i\, , && B \rightarrow -\,B\, , && T_A{}^B \rightarrow -\,T_B{}^A\, ,
&V_{\pm,A} \rightarrow W_{\pm}^A\, , && W_{\pm}^A \rightarrow V_{\pm,A}\, . &&
\end{align}
The $\textbf{6}$ representation can be realized as a set of $2\times 2$ matrices with the Gra\ss{}mann entries acting on a two-component vector of the following form:
\begin{align}
\textbf{x} = \left(\begin{array}{c} x_+ \\ x_- + x_A\Theta^A\end{array}\right)\, .
\end{align}
The $SL(2)$ variables are encoded in the $x_\pm$ components, while the $SU(4)$ part of the algebra acts on $x_A$. Both $x_A$ and $\Theta^A$ are Gra\ss{}mann-odd. The generators that represent the commutation relations above can be chosen in the following way:
\begin{align}\label{2x2alg6}
\mathcal{L}_{-} &= \left(\begin{array}{cc}
0 & -1+\Theta\partial_\Theta \\
0 & 0
\end{array}\right), &
\mathcal{L}_{0} &= \left(\begin{array}{cc}
\frac{1}{2} & 0 \\
0 & \frac{1}{2}\left(-1+\Theta\partial_\Theta\right) 
\end{array}\right), &
\mathcal{L}_{+} &= \left(\begin{array}{cc}
0 & 0 \\
1 & 0
\end{array}\right),\nonumber\\
\mathcal{V}_{-,A} &= \left(\begin{array}{cc}
0 & \partial_{\Theta^A} \\
0 & 0
\end{array}\right), &
\mathcal{B} &= \left(\begin{array}{cc}
\frac{1}{2} & 0 \\
0 & \frac{1}{2}-\frac{1}{4}\Theta\partial_\Theta
\end{array}\right), &
\mathcal{V}_{+,A} &= \left(\begin{array}{cc}
0 & 0 \\
0 & \partial_{\Theta^A}
\end{array}\right),\nonumber\\
\mathcal{W}_{-}^A &= \left(\begin{array}{cc}
0 & 0 \\
0 & \Theta^A\left(-1+\Theta\partial_\Theta\right) 
\end{array}\right), &
\mathcal{T}_{A}{}^B &= \left(\begin{array}{cc}
0 & 0 \\
0 & \Theta^B\partial_{\Theta_A} - \frac{1}{4}\delta_{A}^B\,\Theta\partial_\Theta
\end{array}\right), &
\mathcal{W}_{+}^A &= \left(\begin{array}{cc}
0 & 0 \\
\Theta^A & 0
\end{array}\right),
\end{align}
where $\Theta\partial_\Theta = \Theta^A\partial_{\Theta^A}$. The $\bar{\textbf{6}}$ representation acts on the conjugate Gra\ss{}mann variable $\Omega_A$. The six components of the representation are encoded in the following two-dimensional vector,
\begin{align}
\bar{\textbf{x}} = \left(\begin{array}{c} \bar{x}_+ \\ \bar{x}_- + \bar{x}^A\,\Omega_A\end{array}\right),
\end{align}
which is acted on by the following set of generators
\begin{align}\label{2x2alg6b}
\bar{\mathcal{L}}_{-} &= \left(\begin{array}{cc}
0 & -1+\Omega\partial_\Omega \\
0 & 0
\end{array}\right), 
&\bar{\mathcal{L}}_{0} &= \left(\begin{array}{cc}
\frac{1}{2} & 0 \\
0 & \frac{1}{2}\left(-1+\Omega\partial_\Omega\right) 
\end{array}\right)\, , &
\bar{\mathcal{L}}_{+} &= \left(\begin{array}{cc}
0 & 0 \\
1 & 0
\end{array}\right),\nonumber\\
\bar{\mathcal{V}}_{-,A} &= \left(\begin{array}{cc}
0 & 0 \\
0 & \Omega_A\left(-1+\Omega\partial_\Omega\right)
\end{array}\right), &
\bar{\mathcal{B}} &= \left(\begin{array}{cc}
-\frac{1}{2} & 0 \\
0 & -\frac{1}{2}+\frac{1}{4}\Omega\partial_\Omega
\end{array}\right), &
\bar{\mathcal{V}}_{+,A} &= \left(\begin{array}{cc}
0 & 0 \\
\Omega_A & 0
\end{array}\right)\, ,\nonumber\\
\bar{\mathcal{W}}_{-}^A &= \left(\begin{array}{cc}
0 & \partial_{\Omega_A} \\
0 & 0
\end{array}\right), &
\bar{\mathcal{T}}_{A}{}^B &= \left(\begin{array}{cc}
0 & 0 \\
0 & -\,\Omega_A\partial_{\Omega_B} + \frac{1}{4}\delta_{A}^B\,\Omega\partial_\Omega
\end{array}\right), &
\bar{\mathcal{W}}_{+}^A &= \left(\begin{array}{cc}
0 & 0 \\
0 & \partial_{\Omega_A}
\end{array}\right).
\end{align}
The representations $\textbf{6}$ and $\bar{\textbf{6}}$ can be contracted into an $SL(2|4)$ singlet using the invariant bilinear form $\varepsilon$ defined as
\begin{align}\label{epsilonn}
\varepsilon\left(\bar{\textbf{x}},\textbf{y}\right) = \underbrace{\bar{x}_+y_- - \bar{x}_-y_+}_{SL(2)\text{ singlet}} \,\,+\!\!\! \underbrace{\vphantom{\bar{x}_+y_-}\,\,\bar{x}^Ax_A\,\,\vphantom{a_0}}_{SU(4)\text{ singlet}}\!\!\!, \quad (-1)^{\text{deg}(\mathcal{G})\,\text{deg}(\bar{\textbf{x}})}\,\varepsilon\left(\bar{\mathcal{G}}\cdot \bar{\textbf{x}},\textbf{y}\right) + \varepsilon\left(\bar{\textbf{x}},\mathcal{G}\cdot \textbf{y}\right) = 0\, ,
\end{align}
where $(-1)^{\text{deg}(\mathcal{G})\,\text{deg}(\bar{x})}$ equals $-1$ when both the generator $\mathcal{G}$ and the component of $\bar{\textbf{x}}$ it is acting on are Gra\ss{}mann-odd, and $1$ otherwise.

The explicit form of the small solutions we used to construct the transfer matrices (\ref{Tgen}) depends on the positions of the left and right edges \cite{Sever:2012qp}. In the frame in which these edges are located at zero and infinity, the small solutions take the form 
\beq
s_L = \left(\begin{array}{c} 1 \\ 0 \end{array}\right)\in \bar{\textbf{6}}\, ,\qquad s_R = \left(\begin{array}{c} 0 \\ 1 \end{array}\right)\in \textbf{6}\, ,
\eeq
which, up to irrelevant overall normalization factors, are fixed by demanding
\beq
\bar{\cal L}_-\,s_R = \bar{\cal W}_\pm\,s_R = 0\, ,\qquad {\cal L}_+\,s_R = {\cal V}_\pm\,s_R = 0\, .
\eeq}
The second type of generators that enter the $R$-matrix (\ref{RRbar}) act on the physical space represented by one-particle states of the form
\begin{equation}
 |\sigma,\theta\rangle = \left(x'(\sigma)\right)^{\frac{1+|2-\theta\partial_{\theta}|}{2}}\Phi(x(\sigma),\theta)\, ,
\end{equation}
where $\Phi$ is the superfield (\ref{supermultipletdef}). In this representation, the Gra\ss{}mann-even generators take the following form,
\begin{align}\label{algbb}
&\mathbb{L}_{\pm} = \frac{1}{2}\,e^{\mp 2\sigma}\left(\partial_\sigma \mp 1\mp |2-\theta\partial_\theta|\right)\, ,&& \mathbb{L}_{0} = \frac{1}{2}\,\partial_\sigma\, ,
\\
&\mathbb{B} = \frac{1}{2} - \frac{1}{4}\,\theta\partial_\theta\, , && \mathbb{T}_{A}{}^B = \theta^B\partial_{\theta^A} - \frac{1}{4}\,\delta_{A}^B\,\theta\partial_\theta \nn\, .
\end{align}
The form of the generators in the Gra\ss{}mann-odd sector varies depending on the value of $a$, the eigenvalue of $\theta\partial_\theta=\theta^A\partial_{\theta^A}$, for the component they are acting on. For the first type of generators, one finds
\begin{align}\label{algbb1}
\mathbb{V}_{\pm,A} =\begin{cases}
\frac{1}{2}\,e^{\mp\sigma}\,\partial_{\theta^A}\left(\partial_\sigma \mp 3 \pm \theta\partial_\theta\right) &
\text{ for }a\leq 2\, ,\\
e^{\mp\sigma}\,\partial_{\theta^A}  &
\text{ for }a> 2\, ,
\end{cases}
\end{align}
while for the second type
\begin{align}\label{algbb2}
\mathbb{W}_{\pm}^{A} = \begin{cases}
e^{\mp\sigma}\,\theta^A&\text{ for }a< 2\, ,\\
 \frac{1}{2}\,e^{\mp\sigma}\,\theta^A\left(\partial_\sigma \pm 1 \mp \theta\partial_\theta\right)&\text{ for }a\geq 2\, .
 \end{cases}
 \end{align}
The $n$-particle state transforms in the tensor product of $n$ copies of the $SL(2|4)$ algebra, which act separately on each individual excitation.

The eigenvalue of the Casimir (\ref{Cas}) is the same for $\textbf{6}$, $\bar{\textbf{6}}$ and the physical representations and is equal to $-\frac{3}{4}$.

\section{Square and pentagon transitions and wave-function normalization}\label{SWFappendix}

In this appendix, we consider square and pentagon transitions as operators that map states on the bottom of the corresponding polygon to ones on the top. We then check that the wave functions we constructed lead to  square transitions that are correctly normalized to be equal to the inverse measures of the corresponding two-particle states. Lastly, we introduce the \emph{shadow wave functions}, which can be obtained as images of the regular wave functions under the square transition map. These objects will be instrumental in appendix~\ref{PTappendix}, where we will use them to confirm that our construction leads to correct expressions for the pentagon transitions between two-particle states in the singlet sector.

The square and pentagon transitions can be thought of as operators that transform states on the bottom of the corresponding polygon into ones on the top. We will be using the following universal definition of these objects:
\begin{align}\label{transitiondefff}
\mathcal{X}\cdot\Psi(\boldsymbol{\sigma},\boldsymbol{\theta}) \equiv \langle\boldsymbol{\sigma},\boldsymbol{\theta}|\mathcal{X}|\Psi\rangle = \int d^n\!\boldsymbol{\rho}\,d^{4n}\!\boldsymbol{\zeta}\,\langle\boldsymbol{\sigma},\boldsymbol{\theta}|\mathcal{X}|\boldsymbol{\rho},\boldsymbol{\zeta}\rangle\,\Psi(\boldsymbol{\rho},\boldsymbol{\zeta})\, ,
\end{align}
where $\mathcal{X}\in\{{\mathcal Sq},\mathcal{P}\}$ labels the type of transition in question and $\langle\boldsymbol{\sigma},\boldsymbol{\theta}|\mathcal{X}|\boldsymbol{\rho},\boldsymbol{\zeta}\rangle$ is the corresponding transition in position space, which can be constructed directly from Feynman diagrams. These objects can be obtained as products of one-particle transitions:
\begin{align}
\langle\boldsymbol{\sigma},\boldsymbol{\theta}|\mathcal{X}|\boldsymbol{\rho},\boldsymbol{\zeta}\rangle = \prod\limits_i\langle\sigma_i,\theta_i|\mathcal{X}|\rho_i,\zeta_i\rangle\, .
\end{align}
The one-particle square and pentagon transitions closely resemble the two-particle FF transitions (\ref{1pptr}):
\begin{align}
\langle\sigma,\theta|{\mathcal Sq}|\rho,\zeta\rangle =\,\,& \frac{\theta^{\vphantom{[}1}\theta^{\vphantom{[}2}\theta^{\vphantom{[}3}\theta^{\vphantom{[}4} +\zeta^{\vphantom{[}1}\zeta^{\vphantom{[}2}\zeta^{\vphantom{[}3}\zeta^{\vphantom{[}4}}{\left(e^{\sigma-\rho}+e^{\rho-\sigma}\right)^3} - 2\,\frac{\theta^{[1}\theta^{\vphantom{[}2}\theta^{\vphantom{[}3}\zeta^{4]} - \theta^{[1}\zeta^{\vphantom{[}2}\zeta^{\vphantom{[}3}\zeta^{4]}}{\left(e^{\sigma-\rho}+e^{\rho-\sigma}\right)^2} + 3\,\frac{\theta^{[1}\theta^{\vphantom{[}2}\zeta^{\vphantom{[}3}\zeta^{4]}}{\left(e^{\sigma-\rho}+e^{\rho-\sigma}\right)}\, ,\nonumber\\
\langle\sigma,\theta|\mathcal{P}|\rho,\zeta\rangle =\,\,& \frac{\theta^{\vphantom{[}1}\theta^{\vphantom{[}2}\theta^{\vphantom{[}3}\theta^{\vphantom{[}4} +\zeta^{\vphantom{[}1}\zeta^{\vphantom{[}2}\zeta^{\vphantom{[}3}\zeta^{\vphantom{[}4}}{\left(e^{\sigma-\rho}+e^{\rho-\sigma}+e^{\sigma+\rho}\right)^3} - 2\,\frac{\theta^{[1}\theta^{\vphantom{[}2}\theta^{\vphantom{[}3}\zeta^{4]} - \theta^{[1}\zeta^{\vphantom{[}2}\zeta^{\vphantom{[}3}\zeta^{4]}}{\left(e^{\sigma-\rho}+e^{\rho-\sigma}+e^{\sigma+\rho}\right)^2}\nonumber\\
&+ 3\,\frac{\theta^{[1}\theta^{\vphantom{[}2}\zeta^{\vphantom{[}3}\zeta^{4]}}{\left(e^{\sigma-\rho}+e^{\rho-\sigma}+e^{\sigma+\rho}\right)}\, .
\end{align}
For the specific case of two-particle singlets states, which we are interested in, this general formula for the two-particle transition reduces to the following expression:
\begin{align}\label{2ppsq}
\langle\sigma_1,\sigma_2;\theta_1,\theta_2|{\mathcal Sq}|\rho_1,\rho_2;\zeta_1,\zeta_2\rangle &= \frac{P^1P^2P^3P^4 + Q^1Q^2Q^3Q^4}{\left(e^{\sigma_1-\rho_1}+e^{\rho_1-\sigma_1}\right)^3\left(e^{\sigma_2-\rho_2}+e^{\rho_2-\sigma_2}\right)^3}\nonumber\\
&\phaneq+ \frac{P^{\{1}P^2P^3Q^{4\}}+P^{\{1}Q^2Q^3Q^{4\}}}{\left(e^{\sigma_1-\rho_1}+e^{\rho_1-\sigma_1}\right)^2\left(e^{\sigma_2-\rho_2}+e^{\rho_2-\sigma_2}\right)^2}\nonumber\\
&\phaneq+ \frac{3}{2}\,\frac{P^{\{1}P^2Q^3Q^{4\}}}{\left(e^{\sigma_1-\rho_1}+e^{\rho_1-\sigma_1}\right)\left(e^{\sigma_2-\rho_2}+e^{\rho_2-\sigma_2}\right)}\, .
\end{align}
and an analogous expression for the pentagon transition. Here we introduced
\begin{align}
P^A = \theta_1^A\zeta_2^A\qquad \text{and}\qquad Q^A = \theta_2^A\zeta_1^A\, .
\end{align}
One can see that the two transitions defined in (\ref{transitiondefff}) are not fully independent. In fact, the pentagon-transition operator ${\mathcal P}$ differs from the square-transition operator ${\mathcal Sq}$ by a change of conformal frame on one of the edges. They can therefore be expressed in terms of one another,
\begin{align}\label{SqtoP}
{\mathcal P}\cdot\Psi(\boldsymbol{\sigma},\boldsymbol{\theta}) = \prod\limits_{i=1}^n\frac{1}{\left(1+e^{2\sigma_i}\right)^{\frac{1}{2}+|1-\frac{1}{2}\theta_i\partial_{\theta_i}|}}\,{\mathcal Sq}\cdot\Psi\left(-\,\tfrac{1}{2}\,{\rm log}\left(1+e^{-2\boldsymbol{\sigma}}\right),\boldsymbol{\theta}\right)\, .
\end{align}

The transitions (\ref{transitiondefff}) transfer the wave functions from the bottom of the square to the top, where they can be overlapped with conjugate wave functions using the $SL(2|4)$-invariant scalar product defined as follows:
\begin{align}\label{scalprodd}
\langle\Psi|\Phi\rangle \equiv \int\limits_{\sigma_1<\ldots<\sigma_n} d^n\!\boldsymbol{\sigma}\,d^{4n}\boldsymbol{\theta}\,\Psi^*(\boldsymbol{\sigma},\boldsymbol{\theta})\,\Phi(\boldsymbol{\sigma},\boldsymbol{\theta})\, ,
\end{align}
where in addition to the regular complex conjugation, the $*$ operation interchanges $\theta_1$ and $\theta_2$ and changes the sign of the fermionic components. In other words,
\begin{align}\label{cnjg}
\Psi^* \equiv \left(\Psi^{F\bar{F}}\right)^* \theta_1^{\vphantom{[}1}\theta_1^{\vphantom{[}2}\theta_1^{\vphantom{[}3}\theta_1^{\vphantom{[}4} &- 4\left(\Psi^{\psi\bar{\psi}}\right)^* \theta_1^{[1}\theta_1^{\vphantom{[}2}\theta_1^{\vphantom{[}3}\theta_2^{4]} + 6\left(\Psi^{\phi\bar{\phi}}\right)^* \theta_1^{[1}\theta_1^{\vphantom{[}2}\theta_2^{\vphantom{[}3}\theta_2^{4]}\nonumber\\
&- 4\left(\Psi^{\bar{\psi}\psi}\right)^* \theta_1^{[1}\theta_2^{\vphantom{[}2}\theta_2^{\vphantom{[}3}\theta_2^{4]} + \left(\Psi^{\bar{F}F}\right)^* \theta_2^{\vphantom{[}1}\theta_2^{\vphantom{[}2}\theta_2^{\vphantom{[}3}\theta_2^{\vphantom{[}4}\, .
\end{align}
This definition ensures that the norm (\ref{scalprodd}) is positive definite. The space of wave functions is closed under conjugation, as long as it is extended to include the wave functions with labels $\Phi\bar{\Phi} = \bar{F}F$ and $\Phi\bar{\Phi} = \bar{\psi}\psi$. These additional wave functions are not independent and can be obtained from those with labels $\Phi\bar{\Phi} =F\bar{F}$ and $\Phi\bar{\Phi} =\psi\bar{\psi}$  by a simple permutation of the momenta:
\begin{align}
\Psi_{\bar{\Phi}\Phi}(\sigma_1,\sigma_2;\theta_1,\theta_2|u_1,u_2) \equiv \Psi_{\Phi\bar{\Phi}}(\sigma_1,\sigma_2;\theta_1,\theta_2|u_2,u_1)\, .
\end{align}
Under the conjugation rule (\ref{cnjg}), the extended space of wave functions is mapped onto itself:
\begin{align}
\Psi^*_{\Phi\bar{\Phi}}(\sigma_1,\sigma_2;\theta_1,\theta_2|u_1,u_2) = (-1)^{2\xi_\Phi}\,\Psi_{\bar{\Phi}\Phi}(\sigma_1,\sigma_2;\theta_1,\theta_2|-u_1,-\,u_2)\, .
\end{align}

The wave functions computed in section \ref{wfsectionexplicit} are normalized in such a way that the square transition between them gives the measure of the corresponding two-particle excitation,
\begin{align}\la{phiSqpsi}
&{}_{\Phi\bar{\Phi}}\langle u_1,u_2|{\mathcal Sq}|v_1,v_2\rangle_{\Upsilon\bar{\Upsilon}} \\
&=d_\Phi\,\frac{(2\pi)^2}{\mu_\Phi(u_1)\,\mu_\Phi(u_2)}\,\Big(\delta_{\Phi\Upsilon}\,\delta(u_1-v_1)\,\delta(u_2-v_2) + \delta_{\Phi\bar{\Upsilon}}\,S_{\Phi\bar{\Phi}}(v_1,v_2)\,\delta(u_1-v_2)\,\delta(u_2-v_1)\Big)\, ,\nonumber
\end{align}
where $S_{\Phi\bar{\Phi}}$ are the singlet $S$-matrices (\ref{ssm}) and $\mu_\Phi(u)$ are the measures (\ref{measures}) of the one-particle excitations. Combinatorial factors $d_\Phi$ count the number of fermions and real scalars,
\begin{align}\label{dPhidef}
d_{F} = 1\, ,\qquad d_{\psi} = 4\, ,\qquad d_{\phi} = 6\, .
\end{align}

The wave functions of section \ref{wfsectionexplicit} are orthogonal to each other with respect to the square measure (\ref{phiSqpsi}), but not with respect to the scalar product (\ref{scalprodd}). It is therefore convenient to introduce a new set of wave functions $\eta_{\Phi\bar{\Phi}}$, which we will call \emph{shadow} wave functions,\footnote{This is because the square transition is the $SL(2,{\mathbb R})_\sigma$ shadow transformation 
\cite{SimmonsDuffin:2012uy}.} that are by definition orthogonal to the  wave functions $\Psi_{\Phi\bar{\Phi}}$,
\begin{align}
&{}_{\Phi\bar{\Phi}}\langle \widetilde{u_1,u_2}|v_1,v_2\rangle_{\Upsilon\bar{\Upsilon}} = {}_{\Phi\bar{\Phi}}\langle u_1,u_2|\widetilde{v_1,v_2}\rangle_{\Upsilon\bar{\Upsilon}}\\
&= d_\Phi\,\pi^2\,\Big(\delta_{\Phi\Upsilon}\,\delta(u_1-v_1)\,\delta(u_2-v_2) + \delta_{\Phi\bar{\Upsilon}}\,S_{\Phi\bar{\Phi}}(v_1,v_2)\,\delta(u_1-v_2)\,\delta(u_2-v_1)\Big)\, ,\nonumber
\end{align}
where the tilde on top of the rapidities indicates that the corresponding bra or ket represents the shadow wave function. The shadow wave functions are related to the regular wave functions by the square transition as
\begin{align}\label{Sqdiagg}
{\mathcal Sq}\cdot\Psi_{\Phi\bar{\Phi}}(\sigma_1,\sigma_2;\theta_1,\theta_2|u_1,u_2) = \frac{4}{\mu_\Phi(u_1)\,\mu_{\bar{\Phi}}(u_2)}\,\eta_{\Phi\bar{\Phi}}(\sigma_1,\sigma_2;\theta_1,\theta_2|u_1,u_2)\, .
\end{align}
This relation allows us to derive a simple, numerically testable expression for the two-particle singlet pentagon transition that we present in the next appendix.

When commuted through the square-transition operator ${\mathcal Sq}$, the wave-function transfer matrices $\mathcal{T}_2$ and $\bar{\mathcal{T}}_2$ turn into their state counterparts $T_2$ and $\bar{T}_2$,
\begin{align}
T_2(u)\cdot{\mathcal Sq} = {\mathcal Sq}\cdot \mathcal{T}_2(-u)\, ,\qquad\bar{T}_2(u)\cdot{\mathcal Sq} = {\mathcal Sq}\cdot \bar{\mathcal{T}}_2(-u)\, .
\end{align}
By comparing the definition of the flux-tube state (\ref{wfdeff}) with the definition of the scalar product (\ref{scalprodd}) and recalling that the transfer matrices $\mathcal{T}_2$ and $\bar{\mathcal{T}}_2$, which are diagonalized by the wave functions $\Psi_{\Phi\bar{\Phi}}$, are obtained from the matrices $T_2$ and $\bar{T}_2$ by integration by parts, one can immediately conclude that the conjugates of the shadow wave functions $\eta^*_{\Phi\bar{\Phi}}$ have to diagonalize $T_2$ and $\bar{T}_2$. Since these transfer matrices have no extra boundary terms, one has to simply solve the following equations,
\begin{align}
&T_2(u)\,\eta^*_{\Phi\bar{\Phi}} = \left(u - u_1 - i\xi_\Phi\right)\left(u - u_2 + i\xi_\Phi\right)\eta^*_{\Phi\bar{\Phi}}\, ,\nonumber\\
&\bar{T}_2(u)\,\eta^*_{\Phi\bar{\Phi}} = \left(u - u_1 + i\xi_\Phi\right)\left(u - u_2 - i\xi_\Phi\right)\eta^*_{\Phi\bar{\Phi}}\, ,
\end{align}
or, equivalently,
\begin{align}
&T_2(u)\,\eta_{\Phi\bar{\Phi}} = \left(u + u_1 + i\xi_\Phi\right)\left(u + u_2 - i\xi_\Phi\right)\eta_{\Phi\bar{\Phi}}\, ,\nonumber\\
&\bar{T}_2(u)\,\eta_{\Phi\bar{\Phi}} = \left(u + u_1 - i\xi_\Phi\right)\left(u + u_2 + i\xi_\Phi\right)\eta_{\Phi\bar{\Phi}}\, .
\end{align}
The solution of these equations is rather similar to the one constructed in section \ref{wfsectionexplicit}. In the normalization (\ref{Sqdiagg}), it is given by
\begin{align}
\eta_{\Phi\bar{\Phi}}^{\Upsilon\bar{\Upsilon}} &= \frac{(-1)^{\delta_{\psi\Upsilon}}}{\mathbb{N}_{\Phi\bar{\Phi}}}\,\frac{\mu_\Phi(u_1)\,\mu_{\bar{\Phi}}(u_2)}{8}\Big[e^{2iu_1\sigma_1+2iu_2\sigma_2+2|\xi_\Phi-\xi_\Upsilon|\left(\sigma_1-\sigma_2\right)}\,G^\Upsilon_{\bar{\Phi}}\left(u_1,u_2\Big|e^{2\left(\sigma_1-\sigma_2\right)}\right)\nonumber\\
&\hphantom{=\frac{(-1)^{\delta_{\psi\Upsilon}}}{\mathbb{N}_{\Phi\bar{\Phi}}}\,\Big[}+e^{2iu_2\sigma_1+2iu_1\sigma_2+2|\xi_{\bar{\Phi}}-\xi_\Upsilon|\left(\sigma_1-\sigma_2\right)}\,G^\Upsilon_{{\Phi}}\left(-u_1,-u_2\Big|e^{2\left(\sigma_1-\sigma_2\right)}\right)\Big]\, .
\end{align}
For the sake of brevity we will only present the gluon component function $G^{F}_\Phi$, while the rest of them can be found in the accompanying Mathematica notebook (\texttt{Born level form factor OPE.nb}):
\begin{align}
G^{F}_\Phi(u_1,u_2|r) &\equiv\Gamma\left(\frac{5}{2}+\xi_\Phi+iu_1\right)\Gamma\left(\frac{5}{2}+\xi_\Phi-iu_2\right)\Gamma\left(iu_2-iu_1-2\xi_\Phi\right)\nonumber\\
&\phaneq{}_2F_1\left(\frac{5}{2}+\xi_\Phi+iu_1,\frac{5}{2}+\xi_\Phi-iu_2,1+2\xi_\Phi+iu_1-iu_2\Big|r\right)\, .
\end{align}
In the $\sigma_1\ll\sigma_2$ limit, the shadow wave functions have the same asymptotic behavior as the wave functions of section \ref{wfsectionexplicit}, up to a proportionality coefficient:
\begin{align}\label{shadowasymp}
\eta_{\Phi\bar{\Phi}}(\sigma_1,\theta_1;\sigma_2,\theta_2|u_1,u_2) &\underset{\sigma_1\ll\sigma_2}{\sim} \left(\frac{\lambda_\Phi(u_1)\,\lambda_{\bar{\Phi}}(u_2)}{8}\right)^2\Psi_{\Phi\bar{\Phi}}(\sigma_1,\theta_1;\sigma_2,\theta_2;\sigma_2,\theta_2|u_1,u_2)\, ,
\end{align}
where $\lambda_\Phi(u)$ has been defined in (\ref{lambdadef}).

\section{Singlet pentagon transitions}\label{PTappendix}
So far, the only consistency check of the two-particle singlet wave functions presented in this paper has been  the construction of the singlet $S$-matrices  (\ref{ssm}). This is not a strong check, as it only probes the asymptotic regime of the corresponding wave functions. Therefore, in order to fully demonstrate the validity of the solutions we found, we will use them to construct the two-particle pentagon transition for the singlet states and compare them with results bootstrapped in \cite{Basso:2010in}.

The two-particle pentagon transition has been shown to factorize into the product of one-particle transitions in the following way:
\begin{align}\label{PTansatz}
P_{{\Phi\bar{\Phi}},{\Upsilon\bar{\Upsilon}}}(u_1,u_2|v_1,v_2) = d_\Phi\,d_\Upsilon\,\frac{P_{\Phi\Upsilon}(u_1|v_1)\,P_{\bar{\Phi}\bar{\Upsilon}}(u_2|v_2)\,P_{\Phi\bar{\Upsilon}}(u_1|v_2)\,P_{\bar{\Phi}\Upsilon}(u_2|v_1)}{P_{\Upsilon\bar{\Upsilon}}(v_1|v_2)\,P_{\Phi\bar{\Phi}}(u_2|u_1)}\,\Pi_{{\Phi\bar{\Phi}},{\Upsilon\bar{\Upsilon}}}\, .
\end{align}
Here, $P_{\Phi\Upsilon}(u|v)$ are the one-particle pentagon transitions given by
\begin{align}
P_{\Phi\Upsilon}(u|v) &= \frac{\Gamma\left(iu-iv+|\xi_\Phi-\xi_\Upsilon|\right)}{\Gamma\left(\frac{1}{2}+|\xi_\Phi|+iu\right)\Gamma\left(\frac{1}{2}+|\xi_\Upsilon|-iv\right)}\,\Omega_{\Phi\Upsilon}\, ,
\end{align}
with
\begin{align}
\Omega &=\left(\begin{array}{ccccc}
-\left(u^2+\frac{1}{4}\right)\left(v^2+\frac{1}{4}\right) & -\,i\sqrt{u^2+\frac{1}{4}} & \sqrt{u^2+\frac{1}{4}} & v \sqrt{u^2+\frac{1}{4}} & 1\\
i\sqrt{v^2+\frac{1}{4}} & uv & i\sqrt{u} & 1 & u\sqrt{v^2+\frac{1}{4}}\\
\sqrt{v^2+\frac{1}{4}} & -\,i\sqrt{v} & 1 & -\,i\sqrt{v} & \sqrt{v^2+\frac{1}{4}}\\
u\sqrt{v^2+\frac{1}{4}} & 1 & i\sqrt{u} & uv & i\sqrt{v^2+\frac{1}{4}}\\
1 & v\sqrt{u^2+\frac{1}{4}} & \sqrt{u^2+\frac{1}{4}} & -\,i\sqrt{u^2+\frac{1}{4}} & -\left(u^2+\frac{1}{4}\right)\left(v^2+\frac{1}{4}\right)
\end{array}\right)\, ,
\end{align}
and $\Omega_{\Phi\Upsilon} \equiv \Omega_{3+2\,\xi_\Phi,3+2\,\xi_\Upsilon}$. $\Pi_{{\Phi\bar{\Phi}},{\Upsilon\bar{\Upsilon}}}(u_1,u_2|v_1,v_2)$ is the coupling-independent matrix part of the transition. Since gluon excitations have no $SU(4)_R$ structure, all matrix parts that involve a gluon state inserted on one of the edges of the pentagon are easy to determine,
\begin{equation}
\begin{aligned}
\Pi_{F\bar{F},F\bar{F}}&= \Pi_{F\bar{F},\bar{F}F} = \Pi_{\bar{F}F,F\bar{F}} = \Pi_{\bar{F}F,\bar{F}F} = 1\, ,\\
\Pi_{\psi\bar{\psi},F\bar{F}} &= \Pi_{\psi\bar{\psi},\bar{F}F} = -\,\Pi_{\bar{\psi}\psi,F\bar{F}} = -\,\Pi_{\bar{\psi}\psi,\bar{F}F} = \frac{i}{u_1-u_2-2i}\, ,\\
\Pi_{F\bar{F},\bar{\psi}\psi} &= \Pi_{\bar{F}F,\bar{\psi}\psi} = -\,\Pi_{F\bar{F},\psi\bar{\psi}} = -\,\Pi_{\bar{F}F,\psi\bar{\psi}} = \frac{i}{v_1-v_2+2i}\, ,\\
\Pi_{\phi\bar{\phi},F\bar{F}} & = -\,\frac{1}{(u_1-u_2-i)\,(u_1-u_2-2i)}\, ,\\
\Pi_{F\bar{F},\phi\bar{\phi}} & = -\,\frac{1}{(v_1-v_2+i)\,(v_1-v_2+2i)}\, .
\end{aligned}
\end{equation}
The first non-trivial matrix part for the pentagon transition between two two-particle scalar states has been found in \cite{Basso:2013aha}. Most of the other non-trivial cases have been studied in \cite{Belitsky:2016vyq}, with the exception of the transitions that involve a fermion state on one edge and a scalar one on the other, which is a new result given in (\ref{ffgg}) below.

In the two-scalar case, the matrix part is given by \cite{Basso:2013aha}
\begin{align}
\Pi_{\phi\bar{\phi},\phi\bar{\phi}}\,{}_{i_1i_2}^{j_1j_2} &= \delta_{i_1i_2}\delta^{j_1j_2}\,\pi_1 + \delta_{i_1}^{j_1}\delta_{i_2}^{j_2}\,\pi_2 + \delta_{i_1}^{j_2}\delta_{i_2}^{j_1}\,\pi_3\, ,
\end{align}
where
\begin{align}
\pi_1 &= \frac{(u_1-v_1)\,(u_2-v_2+i)\left((u_1-v_2)\,(u_2-v_1)+i\,(u_2-v_2)-2\right)}{(u_1-u_2-i)\,(u_1-u_2-2i)\,(v_1-v_2+i)\,(v_1-v_2+2i)}\, ,\nonumber\\
\pi_2 &= -\,\frac{(u_1-v_2)\,(u_2-v_1)+i\,(u_2-v_2)-1}{(u_1-u_2-i)\,(v_1-v_2+i)}\, ,\nonumber\\
\pi_3 &= \frac{(u_1-v_1)\,(u_2-v_2+i)}{(u_1-u_2-i)\,(v_1-v_2+i)}\, .
\end{align}
The singlet transition is obtained by contracting the above expressions with $\frac{1}{36}\,\delta^{i_1i_2}\delta_{j_1j_2}$. This results in
\begin{align}
\Pi_{\phi\bar{\phi},\phi\bar{\phi}} &= \pi_1 + \frac{1}{6}\, .
\end{align}
The matrix parts for pairs of fermions $\psi^{A_1}\bar{\psi}_{A_2}$ on the bottom and $\psi^{B_1}\bar{\psi}_{B_2}$ or $\bar{\psi}_{B_2}\psi^{B_1}$ on the top are given by \cite{Belitsky:2016vyq}
\begin{align}
\Pi_{\psi\bar{\psi},\psi\bar{\psi}}\,{}^{A_1B_1}_{A_2B_2} &= \delta^{A_1}_{A_2}\delta^{B_1}_{B_2}\,\theta_1 + \delta^{A_1}_{B_2}\delta^{B_1}_{A_2}\,\theta_2\, ,\\
\Pi_{\psi\bar{\psi},\bar{\psi}\psi}\,{}^{A_1B_1}_{A_2B_2} &= \delta^{A_1}_{A_2}\delta^{B_1}_{B_2}\,\bar{\theta}_1 + \delta^{A_1}_{B_2}\delta^{B_1}_{A_2}\,\bar{\theta}_2\, ,\nonumber
\end{align}
where
\begin{align}
&\theta_1 = -\,\frac{(u_1-v_1)\,(u_2-v_2+i)}{(u_1-u_2-2i)\,(v_1-v_2+2i)}\, ,&&\theta_2 = 1\, ,\\
&\bar{\theta}_1 = \frac{(u_1-v_1-i)\,(u_2-v_2+2i)}{(u_1-u_2-2i)\,(v_1-v_2+2i)}-1\, ,&&\bar{\theta}_2 = 1\, .\nonumber
\end{align}
The singlet contraction is done with $\frac{1}{16}\,\delta^{A_2}_{A_1}\delta^{B_2}_{B_1}$ and leads to
\begin{align}
\Pi_{\psi\bar{\psi},\psi\bar{\psi}} = \Pi_{\bar{\psi}\psi,\bar{\psi}\psi} = \theta_1 + \frac{1}{4}\, ,\qquad\Pi_{\psi\bar{\psi},\bar{\psi}\psi} = \Pi_{\bar{\psi}\psi,\psi\bar{\psi}} = \bar{\theta}_1 + \frac{1}{4}\, .
\end{align}
The last non-trivial matrix part corresponds to the transition between a fermion state and a scalar state,
\begin{align}\label{ffgg}
\Pi_{\phi\bar{\phi},\psi\bar{\psi}} &= -\,\Pi_{\phi\bar{\phi},\bar{\psi}\psi} = \frac{i}{2}\,\frac{(u_1+u_2+i)\,(v_1+v_2)-2\,u_1u_2-2\,v_1v_2-i\,(u_1+u_2)+\frac{5}{2}}{(u_1-u_2-i)\,(u_1-u_2-2i)\,(v_1-v_2+2i)}\, ,\nn\\
\Pi_{\psi\bar{\psi},\phi\bar{\phi}} &= -\,\Pi_{\bar{\psi}\psi,\phi\bar{\phi}} = \frac{i}{2}\,\frac{(u_1+u_2)\,(v_1+v_2-i)-2\,u_1u_2-2\,v_1v_2+i\,(v_1+v_2)+\frac{5}{2}}{(u_1-u_2-2i)\,(v_1-v_2+i)\,(v_1-v_2+2i)}\, .
\end{align}

Note that the matrix parts of the creation amplitudes required for the check of the ansatz \eqref{ansatz} are obtained from the ones above by applying two mirror transformations following~\cite{Belitsky:2016vyq}.

From the OPE perspective, pentagon transitions are obtained as overlaps of two wave functions on opposite edges of the pentagon. We can use (\ref{SqtoP}) to turn the pentagon transition in question into a square transition and then utilize (\ref{Sqdiagg}) to rewrite the result in terms of the shadow wave functions. After a change of integration variables, $\sigma_i\to\frac{1}{2}\,{\rm log}\,\frac{z_i}{1-z_i}$, which is preformed in order to make the domain of integration compact, we arrive at
\begin{align}\label{2pspt}
{}_{\Phi\bar{\Phi}}\langle u_1,&u_2|{\mathcal P}|v_1,v_2\rangle_{\Upsilon\bar{\Upsilon}} = \frac{1}{\mu_\Upsilon(v_1)\,\mu_{\bar{\Upsilon}}(v_2)}\int\limits_{0<z_1<z_2<1}\frac{dz_1dz_2}{z_1z_2}\,d^4\theta_1d^4\theta_2\\
&\times\frac{\Psi^*_{\Phi\bar{\Phi}}\left(\frac{1}{2}\,{\rm log}\,\frac{z_1}{1-z_1},\frac{1}{2}\,{\rm log}\,\frac{z_2}{1-z_2};\theta_1,\theta_2|u_1,u_2\right)}{\left(\left(1-z_1\right)\left(1-z_2\right)\right)^{\frac{1}{2}-|1-\frac{1}{2}\theta_1\partial_{\theta_1}|}}\,\eta_{\Upsilon\bar{\Upsilon}}\left(\frac{1}{2}\,{\rm log}\,z_1,\frac{1}{2}\,{\rm log}\,z_2;\theta_1,\theta_2\big|v_1,v_2\right).\nonumber
\end{align}
This integral can be taken numerically and compared with the predicted factorized expression (\ref{PTansatz}), which gives a complete agreement for all combinations of states on the top and the bottom of the pentagon,
\begin{align}
{}_{\Phi\bar{\Phi}}\langle u_1,u_2|{\mathcal P}|v_1,v_2\rangle_{\Upsilon\bar{\Upsilon}} = P_{{\Upsilon\bar{\Upsilon}},{\Phi\bar{\Phi}}}(v_1,v_2|u_1,u_2)\, .
\end{align}
The full details of this check can be found in the accompanying Mathematica file \texttt{Born level form factor OPE.nb}.

\bibliographystyle{utphys2}
\bibliography{bib}
\end{document}